\documentclass[11pt,a4paper]{article}
\usepackage{times}
\usepackage{fancyheadings}
\usepackage{amsmath}
\usepackage{amssymb}
\usepackage{epsfig}
\usepackage{a4}

\newcommand{\M}{\mathcal{M}}             
\newcommand{\vecp}[1]{{\vec #1}\;\!'}    
\newcommand{\vecpp}[1]{{\vec #1}\;\!''}  
\newcommand{\rcrit}{r_{\!\scriptscriptstyle\rm crit}} 
\newcommand{\rub}{r_{\!\scriptscriptstyle\rm ub}} 
\newcommand{\sss}{\scriptscriptstyle}

\usepackage[hang,stable]{footmisc}
\usepackage{multibib}
\newcites{MATCHING}
         {The matching problem and related techniques}
\newcites{INFLUENCE}
         {Influence of the global cosmological expansion\\ 
          on the local dynamics}
\newcites{EXPLAIN}
         {Tentative explanations of the Pioneer Anomaly with gravity}
\newcites{MEASURE}
         {Reports on the Pioneer Anomaly}
\newcites{OTHERS}
         {Others}

\title{On the influence of the global cosmological expansion\\ 
on the local dynamics in the Solar System}

\author{Matteo Carrera and Domenico Giulini \\
\mbox{}                                     \\
{\small Physikalisches Institut}            \\
{\small Albert-Ludwigs-Universit\"at}       \\
{\small Hermann-Herder-Stra\ss e 3}         \\
{\small D-79104 Freiburg i.~Br.}            \\
{\small Germany}                            \\
\mbox{}                                     \\
{\small carrera@physik.uni-freiburg.de}     \\
{\small giulini@physik.uni-freiburg.de}     
}

\date{}

\begin{document}
\ifx\href\undefined\else\hypersetup{linktocpage=true}\fi

\maketitle

\begin{abstract}
\noindent
All current models of our Universe based on General Relativity  have 
in common that space is presently in a state of expansion. 
In this expository paper we address the question of whether, and 
to what extent, this expansion influences the dynamics on small 
scales (as compared to cosmological ones), particularly in 
our Solar System. Here our reference order of magnitude for 
any effect is given by the apparent anomalous acceleration of 
the Pioneer 10 and 11 spacecrafts, which is of the order 
of $10^{-9}{\rm m/s^2}$. 
We distinguish between \emph{dynamical} and \emph{kinematical} 
effects and critically review the status of both as presented 
in the current literature. We argue that in the Solar 
System dynamical effects can be safely modeled by suitably 
improved Newtonian equations, showing that effects do exist but 
are smaller by many orders of magnitude compared to our Pioneer 
reference. On the other hand, the kinematical effects need a 
proper relativistic treatment and have 
been argued by others to give rise to an additional acceleration of 
the order $Hc$, where $H$ is the Hubble parameter and $c$ is the 
velocity of light. This simple and suggestive expression is 
intriguingly close to the anomalous Pioneer acceleration. 
We reanalyzed this argument and found a discrepancy by a factor 
of $(v/c)^3$, which strongly suppresses the alleged $Hc$--effect 
for the Pioneer spacecrafts by 13 orders of magnitude. We conclude 
with a general discussion which stresses the fundamental importance 
to understand precisely, i.e. within the full dynamical theory 
(General Relativity), the back-reaction effects of local 
inhomogeneities for our interpretation of cosmological data,
a task which is not yet fully accomplished. Finally, a structured 
literature list of more than 80 references gives an overview over 
the relevant publications (known to us).  
\end{abstract}

\newpage


\setcounter{tocdepth}{2}
\tableofcontents

\newpage


\section{Introduction}
\label{sec:Introduction}
The overall theme of this paper concerns the question of whether 
the global cosmological expansion has any influence on the local 
dynamics and kinematics within the Solar System. Despite many efforts 
in the past, this problem is still debated controversially in the 
current literature and hardly anything is written about it in standard 
textbooks. This is rather strange as the question seems to be of 
obvious interest. Hence there is room for speculations that such 
an influence might exist and be detectable with current experimental 
means. It has even be suggested that it be (partly) responsible for 
the apparently anomalous acceleration of the Pioneer spacecrafts, 
the so-called 
`Pioneer-Anomaly'~\cite{Anderson.etal:1998,Anderson.etal:2002a,
Markwardt:2002,Nieto.Turyshev:2004,Turyshev.etal:2005a,%
Turyshev.etal:2005b}, henceforth abbreviated by PA. 

Existing investigations in this direction arrive at partially 
conflicting conclusions. To resolve this issues at a fundamental 
level one may think to start from an investigation of the 
\emph{embedding-problem} for the Solar System. From the viewpoint of 
General Relativity this means to at least find a solution to 
Einstein's field equations that match the gravitational field 
(i.e. the spacetime metric $g_{\mu\nu}$) of the Sun with that of 
the inner-galactic neighbourhood. In a first approximation the 
region exterior to some radius outside the Sun may be modelled by 
a spherically symmetric (around the Sun's center) constant 
dust-like mass distribution (i.e. matching an exterior Schwarzschild
solution to a Friedman-Robertson-Walker solution). Further 
refinements may then take into account the structured nature of 
the cosmological mass distribution as well as a matching of the 
gravitational field of the Galaxy to that of its cosmological 
environment.

Analytical solutions are known for some embedding problems
under various special assumptions concerning symmetries and 
the  matter content of the cosmological environment:
\begin{itemize}
\item[A.]
Spherical symmetry, matching Schwarzschild to a Friedmann-Robertson-Walker
(FRW) universe with pressureless matter and zero cosmological constant 
$\Lambda$. 
\cite{Einstein.Straus:1945,Schuecking:1954}.
\item[B.]
Obvious generalizations of A to many, non-overlapping regions---
so-called `swiss-cheese models'.
\item[C.]
Generalizations of A,B for $\Lambda\ne 0$ \cite{Balbinot.etal:1988}.
\item[D.]
Generalizations of A to spherically symmetric but inhomogeneous 
Lema\^{\i}tre-Tolman-Bondi cosmological backgrounds
\cite{Bonnor:2000b}.
\end{itemize}
In all these approaches, which are based on the original idea of 
Einstein \& Straus~\cite{Einstein.Straus:1945}, the matching is 
between a strict vacuum solution (Schwarzschild) and some cosmological 
model. Hence, within the matching radius, the solution is strictly
of the Schwarzschild type, so that there will be absolutely no 
dynamical influence of cosmological expansion on local dynamical 
processes \emph{by construction}. The only relevant quantity to be 
determined is the matching radius, $r_S$, (sometimes called the 
`Sch\"ucking radius') as a function of the central mass and the 
time. This can be done quite straightforwardly and it turns out 
to be expressible in the following form:
\begin{equation}
\label{eq:MatchingRadius}
\text{Vol}(r_S)\cdot\rho=M\,.
\end{equation} 
Here $\text{Vol}(r_S)$ is the 3-dimensional volume within a sphere of 
radius $r_S$ in the cosmological background geometry and $\rho$ is the 
(spatially constant but time dependent) cosmological mass-density. For example,
for flat or nearly flat geometries we have $\text{Vol}(r_S)=4\pi r_S^3/3$. 
Thus the defining equation for the matching radius has the following 
simple interpretation: if you want to embed a point mass $M$ into a 
cosmological background of mass density $\rho$ you have to 
place it at the centre of an excised ball whose mass (represented by 
the left hand side of (\ref{eq:MatchingRadius})) is just $M$. This is 
just the obvious dynamical condition that the gravitational pull the 
central mass $M$ exerts on the ambient cosmological masses is just the 
same as that of the original homogeneous mass distribution within the 
ball. The deeper reason for why this argument works is that the 
external gravitational field of a spherically symmetric mass 
distribution is independent of its radial distribution, i.e. just 
depending on the total mass. This well known fact from Newton's 
theory remains true in General Relativity, as one readily sees by 
recalling the uniqueness of the exterior Schwarzschild solution. 
We may hence simply think of the cosmological mass inside a 
sphere of radius $r_S$ as being squashed to a point (or black hole) 
at its centre, without affecting the dynamics outside the radius $r_S$. 
Note that for expanding universes the radius $r_S$ expands with 
it, that is, it is co-moving with the cosmological matter.  

At first sight this simple solution may appear as a convincing argument 
against any influence of cosmic expansion on the scale of our Solar 
System, as $r_S$ certainly extends far beyond it; too far in fact\,! 
If $M$ is the mass of the Sun then $r_S$ turns out to be about 
$175\,\mathrm{pc}$, which is more than a factor of 100 larger than the 
distance to our next star and more than factor of 50 larger than the 
average distance of stars in our Galaxy. But this means that the 
Einstein-Straus solution is totally inappropriate as a model for the  
Solar System's neighbourhood. This changes as one goes to larger 
scales, beyond that of galaxy clusters; see 
Section~\ref{sec:SchueckingAtScales}.   

The Einstein-Straus solution may also be criticized on theoretical 
grounds, an obvious one being its dynamical instability. Slight 
perturbations of the matching radius to larger radii will let it 
increase without bound, slight perturbations to smaller radii will let 
it collapse. This can be proven formally (e.g. \cite{Krasinski:1998}, Ch.\,3) 
but it is also rather obvious, since $r_S$ is defined by the equal and 
opposite gravitational pull of the central mass on one side and the 
cosmological mass on the other. Both pulls increase as one moves 
towards their side, so that the equilibrium position must correspond 
to a local maximum of the gravitational potential. Another criticism 
of the Einstein-Straus solution concerns the severe restrictions under 
which it may be generalized to non spherically-symmetric situations; 
see e.g. \cite{Senovilla.Vera:1997,Mena.etal:2002,Mena.etal:2003,Mena.etal:2004}.

We conclude from all this that we cannot expect much useful 
insight, as regards practically relevant dynamical effects 
within the Solar System, from further studies of models based on the 
Einstein-Straus matching idea.\footnote{However, as stressed in 
Section\,\ref{sec:Summary}, the matching problem certainly is  
important on larger scales.} Rather, we shall proceed in the following 
steps: 
\begin{itemize}
\item[I.]
Discuss an improved Newtonian model including a cosmological expansion
term. This we did and the results are given in 
Section~\ref{sec:NewtonianApproach} below. Our discussion 
complements \cite{Cooperstock.etal:1998} which just makes a 
perturbative analysis, thereby missing all orbits which are 
unstable under cosmological expansion (which do exist). In this 
respect it follows a very similar strategy as proposed in the  
recent paper by Price~\cite{Price:2005} (the basic idea of which 
goes back at least to Pachner's work~\cite{Pachner:1963,Pachner:1964}), 
though we think that there are also useful differences. We also 
supply quantitative estimates and clarify that the improved Newtonian 
equations of motion are written in terms of the right coordinates 
(non-rotating and metrically normalized). The purpose of this model 
is to develop a good physical intuition for the qualitative as well 
as quantitative features of any \emph{dynamical} effects involved.
\item[II.]
Eventually the Newtonian model just mentioned has to be understood 
as a limiting case of a genuinely relativistic treatment. For the 
gravitational case this is done in Section\,\ref{sec:McVittie},
where we employ the McVittie metric to model a spherically symmetric 
mass embedded in a $k=0$ FRW universe. The 
geodesic equation is then, in a suitable limit, shown to lead to the 
improved Newtonian model discussed above. The same holds for the 
electromagnetic case, as we show in Section~\ref{sec:DickePeebles}.
There we take a slight detour to also reconsider a classic argument 
by Dicke \& Peebles~\cite{Dicke.Peebles:1964}, which allegedly 
shows the absence of any relevant dynamical effect of global 
expansion. Its original form only involved the dynamical action 
principle together with some simple scaling argument. Since this reference 
is one of the most frequently cited in this field, and since the 
simplicity of the argument (which hardly involves any real analysis) 
is definitely deceptive, we give an independent treatment that makes 
no use of any scaling behaviour of physical quantities other than 
spatial lengths and times. Our treatment also reveals that the 
original argument by Dicke \& Peebles is insufficient to discuss 
leading order effects of cosmological expansion. It is therefore 
also ineffective in its attempt to contradict 
Pachner~\cite{Pachner:1963,Pachner:1964}.    
\item[III.]
Neither the improved Newtonian model nor other general \emph{dynamical} 
arguments make any statement about possible \emph{kinematical} 
effects, i.e. effects in connection with measurements of 
\emph{spatial distances} and \emph{time durations} in a cosmological 
environment whose geometry changes with time. This is an important 
issue since tracking a spacecraft means to map out its `trajectory',
which basically means to determine its simultaneous spatial distance 
to the observer at given observer times. But we know from General Relativity 
that the concepts of `simultaneity' and `spatial distance' are not 
uniquely defined. This fact needs to be taken due care of when 
analytical expressions for trajectories, e.g. solutions to the 
equations of motion in some arbitrarily chosen coordinate system, 
are compared with experimental findings. In those situations it is 
likely that different kinematical notions of simultaneity and distance 
are involved which need to be properly transformed into each other 
before being compared. For example, these transformations can result 
in additional acceleration terms which have been claimed in the 
literature to be directly relevant to the PA; see 
\cite{Palle:2005,Nottale:2003,Rosales:2002,Rosales.Sanchez-Gomez:1998,%
Ranada:2005,Nieto.etal:2005}. We will confirm the existence of 
such effects in principle, but are in essential disagreement 
concerning their relevance in practice. We think that they have been 
overestimated by about 13 orders of magnitude. The details will be 
given in Section~\ref{sec:Rosales}.
\item[IV.]
Finally we made a systematic scan of the literature on the subject. 
The papers found to be relevant are listed in the bibliography at the 
end, which we subdivided into four sections:  . 
\begin{itemize}
\item[1.]
Papers dealing with the proper matching problem in General Relativity.
\item[2.]
Papers dealing generally with the influence of the global 
cosmological expansion on local dynamics, irrespectively of
whether they work within an improved Newtonian setting or in 
full General Relativity. 
\item[3.]
Papers discussing tentative explanations of the PA by means of 
gravity, mostly by referring to kinematical 
effects of space-time measurements in time dependent background 
geometries.
\item[4.]
Measurements of the PA. These were just for our 
own instruction and are listed for completeness. 
\end{itemize}
\end{itemize}

We believe one can give a fair estimation on the irrelevance of the 
\emph{dynamical} effects in question. Given the weakness of the gravitational 
fields involved, estimations by Newtonian methods should give reliable 
figures of orders of magnitude for the motion of ordinary matter. 
\emph{Kinematical} effects based on the equation for light propagation in an 
expanding background also turn out to be negligible, in contrast to some 
claims in the literature, which we think can be straightened out. 
We end by suggesting possible routes for further research.


\section{The Newtonian approach}
\label{sec:NewtonianApproach}
In order to gain intuition we consider a simple bounded system, say 
an atom or a planetary system, immersed in an expanding cosmos. 
We ask for the effects of this expansion on our local system. 
Does our system expand with the cosmos? Does it expand only partially? 
Or does it not expand at all? 

\subsection{The two-body problem in an expanding universe}
Take a two-body problem with a $1/r^2$ attractive force
between them. For simplicity we think of one mass as being much 
smaller than the other one (this is inessential). This can e.g. 
be a system consisting of two galaxies, a star and a planet 
(or spacecraft), or a (classical) atom given by an electron 
orbiting around a proton. We think this system as being immersed 
in an expanding universe and we model the effect of the cosmological 
expansion by adding to the attraction term an extra term coming from 
the Hubble law $\dot{r}=H r$. Here $H:=\dot{a}/a$ is the Hubble 
parameter, $a(t)$ the cosmological scale factor, and $r$ the 
distance -- as measured in the surface of constant cosmological 
time $t$ -- of two objects that follow the Hubble flow (cosmological 
expansion). The acceleration that results from the Hubble law is 
\begin{equation}
\label{eq:cosmological-acc}
\ddot{r}|_{\mathrm{cosm. acc.}} = \frac{\ddot{a}}{a}r \,.
\end{equation}
Note that, in the sense of General Relativity, a body that is 
co-moving with the cosmological expansion is moving on an inertial 
trajectory, i.e. it moves force free. Forces in the Newtonian sense 
are now the cause for \emph{deviations} form the co-moving acceleration 
described by (\ref{eq:cosmological-acc}). This suggests that in Newton's 
law, $m\ddot{\vec r}=\vec F$, we have to replace $\ddot{\vec r}$ by 
$\ddot{\vec r}-(\ddot a/a)\vec r$. This can be justified rigorously by 
using the equation of geodesic deviation in General Relativity. 
In order to do this one must make sure that the Newtonian equations 
of motion are written in appropriate coordinates. That is, they must 
refer to a (locally) non-rotating frame and directly give the 
spatial geodesic distance. This is achieved by using Fermi normal 
coordinates along the worldline of a geodesically moving 
observer---in our case e.g. the Sun or the proton---, as correctly 
emphasized in \cite{Cooperstock.etal:1998}. The equation of geodesic 
deviation in these coordinates now gives the variation of the spatial 
geodesic distance to a neighbouring geodesically moving object---in 
our case e.g. the planet (or spacecraft) or electron. 
It reads\footnote{By construction 
of the coordinates, the Christoffel symbols $\Gamma^\mu_{\alpha\beta}$ 
vanish along the worldline of the first observer. Since this worldline
is geodesic, Fermi-Walker transportation just reduces to parallel 
transportation. This gives a non-rotating reference frame that can be 
physically realized by gyros taken along the worldline.}   
\begin{equation}
\label{eq:GeodDev}
\frac{d^2x^k}{d\tau^2}
+R^k_{\ 0l0 }x^l=0\,.
\end{equation}
Here the $x^k$ are the spatial non-rotating normal coordinates whose 
values directly refer to the proper spatial distance. In these 
coordinates we further have~\cite{Cooperstock.etal:1998}
\begin{equation}
R^k_{\ 0l0 } = - \delta^k_l \, \ddot a/a
\end{equation}
on the worldline of the first observer, where the overdot refers to 
differentiation with respect to the cosmological time, which reduces 
to the eigentime along the observer's worldline.    

Neglecting large velocity effects (i.e. terms quadratic or higher 
order in $v/c$) we can now write down the equation of motion 
for the familiar two-body problem. After specification of a scale 
function $a(t)$, we get two ODEs for the variables $(r,\varphi)$, which 
describe the position\footnote{Recall that `position' refers to 
Fermi normal coordinates, i.e. $r$ is the radial geodesic distance 
to the observer at $r=0$.} of the orbiting body with respect to the 
central one:
\begin{subequations}
\label{eq:impr-N-eqs}
\begin{align}
\label{eq:r-eq}
&\ddot{r} = \frac{L^2}{r^3} - \frac{C}{r^2} + \frac{\ddot{a}}{a}r \\
\label{eq:phi-eq}
&r^2\dot{\varphi}=L \,.
\end{align}
\end{subequations}
These are the $(\ddot a/a)$--improved Newtonian equations of motion
for the two-body problem, where $L$ represents the (conserved) angular 
momentum of the planet (or electron) per unit mass and $C$ the 
strength of the attractive force. In the gravitational case $C=GM$, 
where $M$ is the mass of the central body, and in the electromagnetic 
case $C=|Qq|/4\pi\epsilon_0 m$ (SI-unit), where $Qq$ is the product 
of the two charges, $m$ is the electron mass, and $\epsilon_0$ the 
vacuum permittivity. In Sections~\ref{sec:McVittie} and 
\ref{sec:DickePeebles} we will show how to obtain (\ref{eq:impr-N-eqs}) 
in appropriate limits from the full general relativistic treatments. 

We now wish to study the effect the $\ddot a$ term has on the unperturbed 
Kepler orbits. We first make the obvious remark that this term results 
from the \emph{acceleration} and not just the expansion of the universe. 

We first remark that, in the concrete physical cases 
of interest, the time dependence of this term is negligible to a very 
good approximation. Indeed, putting $f:=\ddot{a}/a$, the relative time 
variation of the coefficient of $r$ in~(\ref{eq:cosmological-acc}) is 
$\dot{f}/f$. For an exponential scale function $a(t)\propto\exp(\lambda t)$ 
($\Lambda$-dominated universe) this vanishes, and for a power law
$a(t)\propto t^\lambda$ (for example matter-, or 
radiation-dominated universes) this is $-2H/\lambda$, and hence of 
the order of the inverse age of the universe. If we consider a planet 
in the Solar System, the relevant time scale of the problem is the 
period of its orbit around the Sun. The relative error in the 
disturbance, when treating the factor $\ddot{a}/a$ as constant during 
an orbit, is hence smaller than $10^{-9}$ for the planets in the 
Solar System. For atoms it is much smaller, of course. Henceforth we 
shall neglect this time-dependence of~(\ref{eq:cosmological-acc}). 

Keeping this in mind we set from now on $\ddot{a}/a=\mathrm{const}=:A$. 
Taking the actual value one can write $A=-q_0 H_0^2$, where the index zero 
means `today' and $q$ stands for the cosmological deceleration 
parameter (defined by $q:=-\ddot a/(H^2a)$). Since the force is time 
independent, we can immediately integrate~(\ref{eq:r-eq}) and get
\begin{equation}\label{eq:r-eq-first-order}
\frac{1}{2}\dot{r}^2 + V(r) = E\,,
\end{equation}
where the effective potential is 
\begin{equation}\label{eq:potential}
V(r)=\frac{L^2}{2r^2}-\frac{C}{r}-\frac{A}{2}r^2 \,.
\end{equation}

\subsection{Specifying the initial-value problem}
\label{sec:SpecifyingIVP}
For (\ref{eq:r-eq-first-order}) and (\ref{eq:phi-eq}) we have to 
specify initial conditions 
$(r,\dot r,\varphi,\dot \varphi)(t_0)=(r_0,v_0,\varphi_0,\omega_0)$ at the 
initial time $t_0$. To study the solutions of the above equation for $r$ one 
has to look at the effective potential. For this purpose it is very convenient 
to introduce a length scale and a time scale that naturally arise in the 
problem. The length scale is defined as the radius at which the 
acceleration due to the cosmological expansion has the same magnitude 
as the gravitational (or electromagnetical) attraction. This happens 
precisely at the critical radius
\begin{equation}\label{eq:r-star}
\rcrit := \left( \frac{C}{|A|} \right)^{1/3} \,.
\end{equation}
For $r<\rcrit$ the gravitational (or electromagnetical) attraction dominates, 
whereas for $r>\rcrit$ the effect of the cosmological expansion is the 
dominant one. 

\subsubsection*{Intermezzo: Expressing the critical radius in terms of
cosmological parameters}
\label{sec:RelCosmPara}
We briefly wish to point out how to express the critical radius in 
terms of the cosmological parameters. For this we write: 
\begin{equation}\label{eq:r-star-grav}
\rcrit = \left( \frac{GM}{|q_0|H_0^2} \right)^{1/3}
    \approx \, \left( \frac{M}{M_\odot} \right)^{1/3} 120\,\mathrm{pc}\,,
\end{equation}
where we have used the current values $q_0=-1/2$ and $h_0=0.7$.
It is interesting to note that in the case of zero cosmological 
constant and pressureless matter we recover the Sch\"ucking gluing 
radius~\cite{Schuecking:1954} (compare (\ref{eq:MatchingRadius})):
\begin{equation}\label{eq:Schuecking-radius}
r_S=\left(\frac{M}{(4/3)\pi \rho_m} \right)^{1/3}\,.
\end{equation}
To see this, just recall that for a pressureless matter we have 
$q_0=(1/2)\Omega_m-\Omega_\Lambda$. Then, for a vanishing cosmological 
constant, and using the definition  $\Omega_m=\rho_m \cdot 8\pi G/(3H_0^2)$, 
one gets the above equations immediately.

In the electromagnetic case, for a proton-electron system,
\begin{equation}\label{eq:r-star-elm}
\rcrit = \left( \frac{|Qq|}{4\pi\epsilon_0 m |q_0|H_0^2} \right)^{1/3}
    \approx \, 30 \, \mathrm{AU}\,,
\end{equation}
which is about as big as the Neptune orbit!

\subsubsection*{Back to the initial-value problem}
The time scale we define is the period with respect to the unperturbed 
Kepler orbit (a solution to the above problem for $A=0$) of 
semi-major axis $r_0$. By Kepler's third law it is given by 
\begin{equation}\label{eq:T-Kepler}
T_K := 2\pi \left( \frac{r_0^3}{C} \right)^{1/2} \,.
\end{equation}
It is convenient to introduce two dimensionless parameters which 
essentially encode the initial conditions $(r_0,\omega_0)$.
\begin{alignat}{3}
\label{eq:def-lambda}
&\lambda &&\,:=\,\quad\left( \frac{\omega_0}{2\pi/T_K} \right)^2
         &&\,=\,\frac{L^2}{C r_0}\,, \\
\label{eq:def-alpha}
&\alpha  &&\,:=\,\mathrm{sign}(A)\left( \frac{r_0}{\rcrit} \right)^3
         &&\,=\, A\frac{r_0^3}{C} \, .
\end{alignat}
For close to Keplerian orbits $\lambda$ is close to one. For reasonably 
sized orbits $\alpha$ is close to zero. For example, in the Solar 
System, where $r_0 < 100$ AU, one has $|\alpha| < 10^{-16}$. For an atom 
whose radius is smaller than $10^4$ Bohr-radii we have $|\alpha|<10^{-57}$.

Defining 
\begin{equation}\label{eq:def-x}
x(t):=r(t)/r_0\,,
\end{equation}
equations~(\ref{eq:r-eq-first-order}) and~(\ref{eq:phi-eq}) can now be 
written as
\begin{align}
\label{eq:x-eq}
&\frac{1}{2}\dot{x}^2 + (2\pi/T_K)^2 \, v_{\lambda,\alpha}(x) = e \\
\label{eq:phi-x-eq}
&x^2\dot{\varphi} = \omega_0\,,
\end{align}
where $e:=E/r_0^2$ now plays the r\^ole of the energy-constant and 
where the reduced 2-parameter effective potential $v_{\lambda,\alpha}$ 
is given by  
\begin{equation}
\label{eq:potential-x}
v_{\lambda,\alpha}(x) := \frac{\lambda}{2x^2}
                        -\frac{1}{x}
                        -\frac{\alpha}{2} x^2\,.
\end{equation}
The initial conditions now read
\begin{equation}
\label{eq:InitCond}
(x,\dot{x},\varphi,\dot{\varphi})(t_0)=(1,v_0/r_0,\varphi_0,\omega_0)\,.
\end{equation}
The point of introducing the dimensionless variables is that the 
three initial parameters $(L,C,A)$ of the effective potential 
could be reduced to two: $\lambda$ and $\alpha)$. This will be 
convenient in the discussion of the potential.

\subsection{Discussion of the reduced effective potential}
\label{sec:DiscEffPot}
%
\begin{figure}[ht]
\centering
\includegraphics{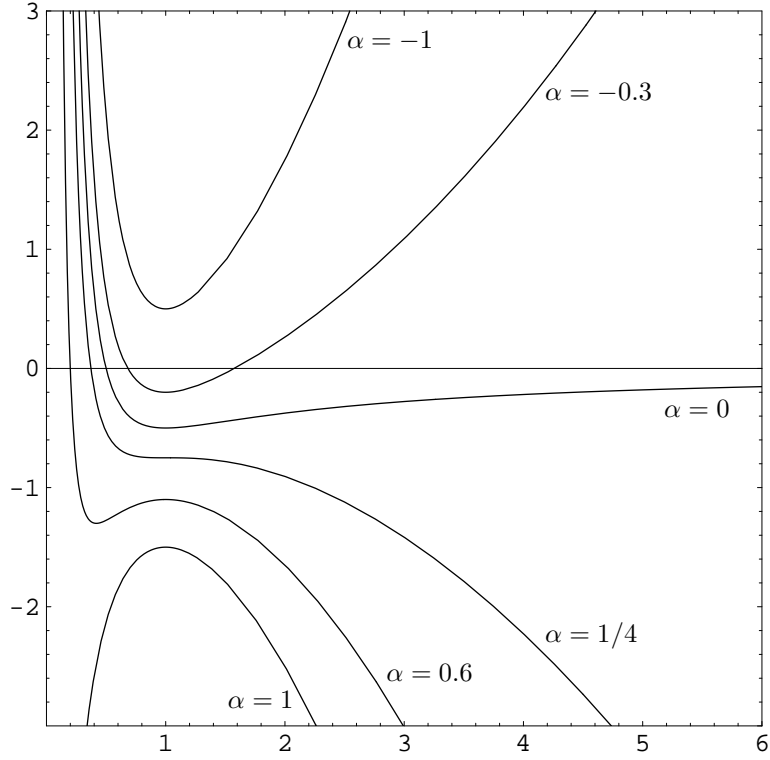}
\put(-160,270){{\small $\alpha = -1   $ }}
\put( -85,250){{\small $\alpha = -0.3 $ }}
\put( -40,130){{\small $\alpha =  0   $ }}
\put( -85, 45){{\small $\alpha =  1/4 $ }}
\put(-145, 30){{\small $\alpha =  0.6 $ }}
\put(-205, 20){{\small $\alpha =  1   $ }}
\caption{The figure shows the effective potential $v_{\lambda,\alpha}$,
for circular orbits, where $\lambda=1-\alpha$, for some values of $\alpha$. 
The initial conditions are $x=1$ and $\dot x=0$ (see~(\ref{eq:def-x})).
At $x=1$ the potential has an extremum, which for $\alpha<1/4$ is a 
local minimum corresponding to stable circular orbits. For 
$1/4 \leq \alpha<1$ these become unstable.}
\label{fig:potential}
\end{figure}

Circular orbits correspond to extrema of the effective 
potential~(\ref{eq:potential}). Expressed in terms of the 
dimensionless variables this is equivalent to 
$v'_{\lambda,\alpha}(1) = -\lambda + 1 - \alpha = 0$.
By its very definition~(\ref{eq:def-lambda}), $\lambda$ is always 
nonnegative, implying $\alpha \leq 1$. For negative $\alpha$ 
(decelerating case) this is always satisfied. On the contrary, for 
positive $\alpha$ (accelerating case), this implies the existence 
of a critical radius 
\begin{equation}\label{eq:r0-max-circle}
r_0 \leq \rcrit
\end{equation}
beyond which no circular orbit exists.

These orbits are stable if the considered extremum is a true minimum, 
i.e. if the second derivative of the potential evaluated at the 
critical value is positive. Now, 
$v''_{\lambda,\alpha}(1)=3\lambda-2-\alpha=1-4\alpha$, showing stability 
for $\alpha < 1/4$ and instability for $\alpha \geq 1/4$. 

Expressing this in physical quantities we can summarize the 
situation as follows: in the decelerating case (i.e. for negative 
$\alpha$ or, equivalently, for negative $A$) stable circular orbits 
exist for every radius $r_0$; one just has to increase the angular 
velocity according to (\ref{eq:omega-circle}). On the contrary, in 
the  accelerating case (i.e. for positive $\alpha$, or, equivalently, 
for positive $A$), we have three regions:
\begin{itemize}
\item $r_0 < \rub := (1/4)^{1/3} \rcrit 
      \approx 0.63 \,\rcrit$, where circular orbits exist and are  
      stable.\footnote{`ub' stands for `upper bound for stable circular 
      orbits'}
\item $\rub \leq r_0 \leq \rcrit$, where
      circular orbits exist but are unstable.
\item $r_0 > \rcrit$, where no circular orbits exist.
\end{itemize}
Generally, there exist no bounded orbits that extend beyond the 
critical radius $\rcrit$, the reason being simply that there is no 
$r>\rcrit$ where $V'(r)>0$. Bigger systems will just be slowly 
pulled apart by the cosmological acceleration and approximately 
move with the Hubble flow at later times.\footnote{This genuine 
non-perturbative behaviour was not seen in the perturbation analysis 
performed in \cite{Cooperstock.etal:1998}.}  

Turning back to the case of circular orbits, we now express the 
condition for an extrema derived above, $\lambda=1-\alpha$, in 
terms of the physical quantities, which leads to 
\begin{equation}\label{eq:omega-circle}
\omega_0=(2\pi/T_K)\sqrt{1-\mathrm{sign}(A)(r_0/\rcrit)^3} \,.
\end{equation}
This equation says that, in order to get a circular orbit, our planet, 
or electron, must have a smaller or bigger angular velocity according 
to the universe expanding in an accelerating or decelerating fashion
respectively. This is just what one would expect, since the effect of 
a cosmological `pulling apart' or `pushing together' must be 
compensated by a smaller or larger centrifugal forces respectively,
as compared to the Keplerian case. Equation~(\ref{eq:omega-circle}) 
represents a modification of the third Kepler law due to the 
cosmological expansion. In principle this is measurable, but it is 
an effect of order $(r_0/\rcrit)^3$ and hence very small indeed; 
e.g. smaller than $10^{-17}$ for a planet in the Solar System.

Instead of adjusting the initial angular velocity as 
in~(\ref{eq:omega-circle}), we can ask how one has to modify 
$r_0$ in order to get a circular orbit with the angular velocity 
$\omega_0=2\pi/T_K$. This is equivalent to searching the minimum 
of the effective potential~(\ref{eq:potential-x}) for $\lambda=1$. 
This condition leads to the fourth order equation 
$\alpha x^4 - x + 1 = 0$ with respect to $x$. Its solutions can be 
exactly written down using Ferrari's formula, though this is not 
illuminating. For our purposes it is more convenient to solve it 
approximatively, treating $\alpha$ as a small perturbation. 
Inserting the ansatz $x_{\mathrm{min}}=c_0+c_1\alpha+O(\alpha^2)$ we get
$c_0=c_1=1$. This is really a minimum since 
$v''_{1,\alpha}(x_{\mathrm{min}})=1+O(\alpha)>0$. Hence we have
\begin{equation}\label{eq:r-min}
r_{\!\scriptscriptstyle\rm min} = r_0
  \left( 1 + \mathrm{sign}(A)\left( \frac{r_0}{\rcrit} \right)^3 
           + O\Big( (r_0/\rcrit)^6 \Big) 
  \right)
\end{equation}
This tells us that in the accelerating (decelerating) case the 
radii of the circular orbits with $\omega_0=2\pi/T_K$ becomes bigger 
(smaller), again according to expectation. As an example, the deviation 
in the radius for an hypothetical spacecraft orbiting around the Sun 
at 100~AU would be just of the order of 1~mm. Since it grows with the 
fourth power of the distance, the deviation at 1000~AU would be  
be of the order of 10 meters.


\section[Fully relativistic treatment for gravitationally-bounded systems:\\ The McVittie model]{Fully relativistic treatment for gravitationally-bounded systems: The McVittie model}
\label{sec:McVittie}
In physics we are hardly ever in the position to mathematically 
rigorously model physically realistic scenarios. Usually we are 
at best either able to provide approximate solutions for realistic 
models or exact solutions for approximate models, and in most cases 
approximations are made on both sides. The art of physics then 
precisely consists in finding the right mixture in each given case.
However, in this process our intuition usually strongly rests on 
the existence of at least some `nearby' exact solutions. 
Accordingly, in this section we seek to find exact solutions in 
General Relativity that, with some degree of physical approximation, 
model a spherically symmetric body immersed in an expanding 
universe.

There are basically two ways to proceed, which could be described 
as the `gluing'- and the `melting'-way respectively. In the first and 
simpler approach one constructs a new solution to Einstein's equations 
by suitably gluing together two known solutions, one corresponding to 
the star (i.e. Schwarzschild, if the star has negligible angular momentum), 
the other to a homogeneous universe (i.e. FRW). The resulting spacetime 
is then divided into two distinct regions, whose interiors are locally 
isometric to the original solutions. The Einstein-Straus-Sch\"ucking vacuole%
~\cite{Einstein.Straus:1945,Schuecking:1954} clearly belongs to this 
class, as well as its generalizations~\cite{Balbinot.etal:1988,Bonnor:2000b}.
The advantage of this gluing-approach is its relative analytic simplicity, 
since the solutions to be matched are already known. Einstein's equations 
merely reduce to the junction conditions along their common seam. 
Its disadvantage is that this gluing only works under those very 
special conditions which allow the glued solutions to locally persist 
\emph{exactly}, and these conditions are likely to be physically unrealistic. 

In the second approach one considers genuinely new solutions of 
Einstein's equations which only approximately resemble the spacetimes 
of an isolated star or a homogeneous universe at small and large 
spatial distances respectively. This `melting-together' is the more 
flexible approach which therefore allows to model physically more 
realistic situations. Needless to say that it also tends to be 
analytically more complicated and that a physical interpretation is
often not at all obvious. The solutions of McVittie~\cite{McVittie:1933} 
and also that of Gautreau~\cite{Gautreau:1984b} fall under this class. 

Our goal is to accurately model the Solar System in the currently 
expanding spatially-flat universe. We already argued in 
Section~\ref{sec:Introduction} that the Einstein-Straus-Sch\"ucking 
vacuole model unfortunately does not apply to the Solar System since 
the matching radius would be much too big (see also 
Appendix~\ref{sec:SchueckingAtScales}). The model of 
Gautreau~\cite{Gautreau:1984b} is much harder to judge. 
Its physical and mathematical assumptions are rather implicit and not 
easy to interpret as regards their suitability for the problem at hand. 
For example, it assumes the cosmic matter to move geodesically 
outside the central body, but at the same time also assumes an equation 
of state in the form $p(\rho)$. For $p\not=0$ this seems contradictory 
in a genuinely non-homogeneous situation since pressure gradients will 
necessarily result in deviations from geodesic motions (see 
\cite{vdBergh.Wils:1984}). Assuming $p=0$ (which implies a motion 
of cosmic matter with non-vanishing shear) Gautreau finds 
in~\cite{Gautreau:1984b} 
that orbits will spiral into the central mass simply because there 
is a net influx of cosmic matter and hence an attracting source 
of increasing strength. This is not really the kind of effect we 
are interested in here. 

Among the models discussed in the literature the one that is best 
understood as regards its analytical structure as well as its physical 
assumptions is that of McVittie~\cite{McVittie:1933} for $k=0$. 
It is therefore, in our opinion, the natural candidate to consider 
first when modeling systems like the Solar System in our expanding 
Universe. As emphasized above, this is not to say that this model 
is to be considered realistic in all its detailed aspects, but at 
least we do have some fairly good control over the assumptions it is 
based on thanks to the carefully analysis by 
Nolan~\cite{Nolan:1998,Nolan:1999a,Nolan:1999b}.
For example, there is a somewhat unrealistic behaviour of the McVittie
spacetime and the matter in it near the singularity at $r=m/2$, 
as discussed below. But at larger radial distances the `flat' 
(i.e. $k=0$) McVittie model may well give a useful description of 
the exterior region of a central object in an expanding spatially-flat 
universe, at least in the region where the radius is much larger 
(in geometric units) than the central mass (to be defined below). 
For planetary motion and spacecraft navigation in the Solar System 
this is certainly the case, since the ratio of the central mass to 
the orbital radius is of the order of $1.5\,{\rm km}/1{\rm AU}=10^{-8}$.




In this section we briefly review the McVittie spacetime and look at 
the geodesic equation in it, showing that it reduces to~(\ref{eq:impr-N-eqs})
in an appropriate weak-field and slow-motion limit. This provides another
more solid justification for the Newtonian approach we carried out 
in Section~\ref{sec:NewtonianApproach}. 

The flat McVittie model (from now on to be simply referred to as 
\emph{the} McVittie model) is characterized by two inputs:
First, one makes the following ansatz for the metric which 
represents an obvious attempt to melt together the Schwarzschild 
metric (in spatially isotropic coordinates) with the spatially flat 
FRW metric~(\ref{eq:FlatFRW1}):
\begin{equation}
\label{eq:McVittieAnsatz}
g = \left( \frac{1-m(t)/2r}{1+m(t)/2r} \right)^2 dt^2
   -\left( 1+\frac{m(t)}{2r} \right)^4 a^2(t)\ (dr^2 + r^2 d\Omega^2) \,.
\end{equation}
Here $d\Omega^2=d\theta^2+\sin^2\theta\,d\varphi^2$ and the two 
time-dependent functions $m$ and $a$ are to be determined.
It is rotationally symmetric with the spheres of constant radius being 
the orbits of the rotation group. Second, it is assumed that the 
ideal fluid (with isotropic pressure) representing the cosmological 
matter moves along integral curves of the vector field 
$\partial/\partial t$. Note that this vector field is \emph{not}
geodesic (unlike in the Gautreau model). Moreover, this model contains 
the implicit assumption that the fluid motion is shearless 
(cf. Chapter\,16 of~\cite{Kramer.etal:2003}).

Einstein's equations together with the equation of state
determine the four functions $m(t), a(t), \rho(t,r)$, and $p(t,r)$.
The former are equivalent to 
\begin{subequations}\label{eq:McV-Einstein}
\begin{alignat}{2}
\label{eq:McV-Einstein-1}
&(a \, m)\!\dot{\phantom{I}}\! &&\,=\, 0 \,,\\
\label{eq:McV-Einstein-2}
&8\pi \rho       &&\,=\, 3 \left( \frac{\dot a}{a} \right)^2 \,,\\
\label{eq:McV-Einstein-3}
&8\pi p          &&\,=\, - 3 \left( \frac{\dot a}{a} \right)^2
- 2 \left( \frac{\dot{a}}{a} \right) \!\!\!\!\dot{\phantom{\frac{I}{I}}}
 \left( \frac{1+m/2r}{1-m/2r} \right) \,.
\end{alignat}
\end{subequations}
Here we already used the first one in order to express the derivatives 
of $m$ in terms of $a$ and its derivatives. The first equation can be 
immediately integrated:
\begin{equation}
\label{eq:mIntegration}
m(t) = \frac{m_0}{a(t)} \,,
\end{equation}
where $m_0$ is an integration constant. Below we will show that this 
integration constant is to be interpreted as the mass of the central
particle. We will call the metric~(\ref{eq:McVittieAnsatz}) together
with condition~(\ref{eq:mIntegration}) the \emph{McVittie metric}.

We note that the equation of state must be necessarily space dependent. 
This follows directly from the equations~(\ref{eq:McV-Einstein-2}) 
and~(\ref{eq:McV-Einstein-3}), which imply that the density depends on 
the time coordinate only whereas the pressure depends on both time and 
space coordinates. Formally the system (\ref{eq:McV-Einstein}) can be 
looked upon in two ways: either one prescribes an equation of state 
and deduces from (\ref{eq:McV-Einstein-2}), (\ref{eq:McV-Einstein-3}), 
and~(\ref{eq:mIntegration}) a second-order differential equation for 
the scale factor $a$, or one specifies $a(t)$ and deduces the 
matter density, the pressure, and hence the equation of state. 

As special cases of (\ref{eq:McV-Einstein}) we remark that if either 
$a$ or $m$ are time independent (\ref{eq:McV-Einstein-1}) implies  
that both must be time independent. This, in turn, implies that density 
and pressure vanish everywhere, resulting in the Schwarzschild solution 
in spatially isotropic coordinates. If we choose the equation of state 
to be that of pressureless dust, i.e. $p=0$, we get either the 
Schwarzschild solution or the dust-filled FRW universe. This follows  
from (\ref{eq:McV-Einstein-3}), where we must distinguish between two 
cases: $\dot{a}/a$ can only be constant if it is zero, 
hence resulting in the Schwarzschild solution. If $\dot{a}/a$ is not 
constant, equation~(\ref{eq:McV-Einstein-3}) (with~(\ref{eq:mIntegration})) 
implies, after a partial differentiation with respect to $r$, 
that $m_0=0$. This gives the homogeneous and isotropic dust-filled 
FRW universe. Another possible equation of state is that of a 
cosmological-constant. This choice implies constancy of $\dot{a}/a$ 
and hence that the second term on the left hand side 
of~(\ref{eq:McV-Einstein-3}) vanishes. In this way one recovers 
the Schwarzschild-de~Sitter metric in spatially isotropic coordinates.

Finally we also mention some critical aspects of the McVittie 
model. In fact, unless $\dot{a}/a$ is constant, it has a singularity at 
$r=m/2$ where the pressure as well as some curvature invariants 
diverge. The former can be immediately seen from~(\ref{eq:McV-Einstein-3}). 
This is clearly a result of the assumption that the fluid moves along 
the integral curves of $\partial/\partial t$, which become lightlike 
in the limit as $r$ tends to $m/2$. Their acceleration is given by 
the gradient of the pressure, which diverges in that limit.
For a study of the singularity at $r=m/2$ 
see~\cite{Nolan:1999a,Nolan:1999b}. 

A related question concerns the global behaviour of the McVittie 
metric. Each hypersurface of constant time $t$ is a complete Riemannian 
manifold which besides the rotational symmetry admits a discrete 
isometry, given in $(r,\theta,\varphi)$ coordinates by 
\begin{equation}
\label{eq:McVittieIsometry}
\phi(r,\theta,\varphi)=
\bigl([m_0/2a(t)]^2\,r^{-1}\,,\,\theta\,,\,\varphi\bigr)\,.
\end{equation} 
It corresponds to a reflection at the 2-sphere $r=(m_0/2a(t))$ and 
shows that the hypersurfaces of constant $t$ can be thought of 
as two isometric asymptotically-flat pieces joined together at the 
totally geodesic (being a fixed-point set of an isometry) 2-sphere 
$r=m_0/2a(t)$, which is minimal. Except for the time-dependent 
factor $a(t)$, this is just like for the slices of constant $t$
in the Schwarzschild metric (the difference being that 
(\ref{eq:McVittieIsometry}) does not extend to an isometry of the 
spacetime metric unless $\dot a=0$). This means that the McVittie 
metric cannot literally be interpreted as corresponding to a 
point particle sitting at $r=0$ ($r=0$ is in infinite metric distance) 
in a flat FRW universe, just like the Schwarzschild metric does not 
correspond to a point particle sitting at $r=0$ in Minkowski space. 
Unfortunately, McVittie seems to have interpreted his solution
in this fashion~\cite{McVittie:1933} which even until recently 
gave rise to some confusion in the literature (e.g. 
\cite{Gautreau:1984b,Sussman:1988,Ferraris.etal:1996}). 
A clarification was given by Nolan~\cite{Nolan:1999a}. Another 
important issue is whether the cosmological matter satisfies
some energy condition. For a discussion about this topic we refer 
to~\cite{Nolan:1998,Nolan:1999a}.

\subsection{Interpretation of the McVittie metric}
\label{sec:McVittieInterpretation}
We shall now present some arguments which justify calling 
McVittie's metric a model for a localized mass immersed in a 
flat FRW background. Here we basically follow~\cite{Nolan:1998}.
As is well known, it is generally not possible in General 
Relativity to assign a definite mass (or energy) to a local
bounded region of space (quasi-local mass). Physically sensible 
definitions of such a concept of quasi-local mass exist only in 
favourable and special circumstances, one of them being spherical 
symmetry. In this case the so-called Misner-Sharp energy is often 
employed (e.g. ~\cite{Misner.Sharp:1964,Cahill.McVittie:1970a,%
Cahill.McVittie:1970b,Burnett:1991,Hayward:1996,Hayward:1998}). 
It allows to assign an energy content to the interior region of any 
two-sphere of symmetry (i.e. an orbit of $SO(3)$). For the McVittie 
metric ((\ref{eq:McVittieAnsatz}) with~(\ref{eq:mIntegration})) 
the Misner-Sharp energy takes the simple and intuitively appealing 
form:
\begin{equation}
\label{eq:MSEnergyOfMcV}
E_{\rm{\scriptscriptstyle MS}}(g;R,t)=\frac{4}{3}\pi R^3 \rho(t) + m_0 \,
\end{equation}
where henceforth we denote by $R$ the `areal radius'
defined by~(\ref{eq:McVittieArealRadius}). This shows that the total 
energy is given by the sum of the cosmological matter contribution and
the central mass, where the mass of the central object is given by  
$m_0$.

Another useful definition of quasi-local mass is that of 
Hawking~\cite{Hawking:1968}. According to this definition the 
energy contained in the region enclosed by a spatial two-sphere 
$S$ is given by a surface integral over $S$, whose integrand is 
essentially the sum of certain distinguished components of the 
Ricci and Weyl tensors, representing the contributions of matter 
and the gravitational field respectively. Applied to the McVittie 
metric the latter takes the value $m_0$ for any 2-sphere outside 
of and enclosing $R=2m_0$. 

\subsection{Motion of a test particle in the McVittie spacetime}
We are interested in the motion of a test particle (idealizing a 
planet or a spacecraft) in McVittie's spacetime. In \cite{McVittie:1933} 
McVittie concluded within a slow-motion and weak-field approximation
that Keplerian orbits do not expand as measured with the `cosmological 
geodesic radius' $r_*=a(t)r$. Later Pachner~\cite{Pachner:1963} and 
Noerdlinger \& Petrosian~\cite{Noerdlinger.Petrosian:1971} argued for 
the presence of the acceleration term~(\ref{eq:cosmological-acc}) 
proportional to $\ddot{a}/a$ within this approximation scheme, hence 
arriving at~(\ref{eq:r-eq}). In the following we shall show how to 
arrive at (\ref{eq:r-eq}) from the exact geodesic equation of the 
McVittie metric by making clear the approximations involved. 
In order to compare our calculation with similar ones in the recent 
literature (i.e. \cite{Bolen.etal:2001,Baker:2002})\footnote{The 
paper~\cite{Bolen.etal:2001} contains a derivation of the effect 
of cosmological expansion on the periastron precession and 
eccentricity change in the case where the Hubble parameter 
$H:=\dot{a}/a$ is constant.} we will work with the 
so-called `areal radius'. It corresponds to a function that can 
be geometrically characterized on any spherically symmetric spacetime 
by taking the square root of the area of the $SO(3)$-orbit through the 
considered point divided by $4\pi$. Hence it is the same as 
the square root of the modulus of the coefficient of the angular 
part of the metric. For the McVittie metric this reads 
\begin{equation}
\label{eq:McVittieArealRadius}
R(t,r)=\left( 1+\frac{m_0}{2a(t)r} \right)^2 a(t)\,r \,.
\end{equation}
Note that for fixed $t$ the map $r\mapsto R(t,r)$ is 2-to-1 and that 
$R\geq 2m_0$, where $R=2m_0$ corresponds to $r=m_0/2a$. Hence we 
restrict the coordinate transformation (\ref{eq:McVittieArealRadius}) 
to the region $r> m_0/2a$ where it becomes a diffeomorphism onto the 
region $R> 2m_0$.  

Reintroducing factors of $c$, McVittie's metric assumes the (non-diagonal) 
form in the region $R>2m_0$ (i.e. $r>m_0/2a(t)$)
\begin{equation}
\label{eq:McVittieMetric2}
g = \left( f(R) - \left( \frac{H(t)R}{c} \right)^2 \right)c^2 dt^2
   +\frac{2(H(t)R/c)}{\sqrt{f(R)}} c\,dt\,dR
   -\frac{dR^2}{f(R)}
   - R^2 d\Omega^2 \,,
\end{equation}
where we put 
\begin{alignat}{2}
\label{eq:Def-f}
&f(R) &&:= 1-\frac{2m_0}{R} \,,\\
\label{eq:Def-H}
&H(t) &&:= \frac{\dot a}{a}(t) \,.
\end{alignat}
The region $R<2m_0$ was investigated in \cite{Nolan:1999b}.

The equations for a timelike geodesic (i.e. parameterized with respect 
to eigentime) $\tau \mapsto z^\mu(\tau)$ with $g(\dot z,\dot z)=c^2$ 
follows via variational principle from the Lagrangian 
$\mathcal{L}(z,\dot z)=(1/2)g_{\mu\nu}(z){\dot z^\mu}{\dot z^\nu}$.
Spherical symmetry implies conservation of angular momentum. 
Hence we may choose the particle orbit to lie in the 
equatorial plane $\theta=\pi/2$. The constant modulus of angular 
momentum is  
\begin{equation}
\label{eq:ConsAngMom}
R^2\dot{\varphi}=L \,.
\end{equation}
The remaining two equations are then coupled second-order ODEs for 
$t(\tau)$ and $R(\tau)$. However, we may replace the first one by its 
first integral that results from $g(\dot z,\dot z)=c^2$: 
\begin{equation}
\label{eq:TimeEquation}
\left( f(R) - \left( \frac{H(t)R}{c} \right)^2 \right)c^2 \dot t^2
   +\frac{2(H(t)R/c)}{\sqrt{f(R)}} c\,\dot t\,\dot R
   -\frac{\dot R^2}{f(R)}
   - (L/R)^2 =c^2\,.
\end{equation}

The remaining radial equation is given by 
%
%
\begin{subequations}\label{eq:McV-R-sec}
\begin{alignat}{1}
\ddot{R} \;
&-\left( f(R)-\left(\frac{H(t)R}{c}\right)^2 \right)\frac{L^2}{R^3}
\label{eq:McV-R-sec-1}\\
&+\frac{m_0\,c^2}{R^2}f(R)\,\dot{t}^2
\label{eq:McV-R-sec-2}\\
&-R\Bigg(
       \dot{H}(t)\,f(R)^\frac{1}{2}+
       H(t)^2\bigg( 1-\frac{m_0}{R}-\left(\frac{H(t)R}{c}\right)^2 \bigg)
    \Bigg)\,\dot{t}^2
\label{eq:McV-Rsec-3}\\
&-\left( \frac{m_0}{R}-\left(\frac{H(t)R}{c}\right)^2 \right)
  f(R)^{-1}\,\frac{\dot R}{R}^2
\label{eq:McV-R-sec-4}\\
&+2\left( \frac{m_0}{R}-\left(\frac{H(t)R}{c}\right)^2 \right)
   f(R)^{-\frac{1}{2}}\, c\,H(t)\,(\dot{R}/c)\,\dot{t} \;\; = \; 0 \,,
\label{eq:McV-R-sec-5}
\end{alignat}
\end{subequations}
Recall that $m_0=GM/c^2$, where $M$ is the mass of the central star
(the Sun in our case) in standard units (kg).

Equations (\ref{eq:TimeEquation},\ref{eq:McV-R-sec}) are exact.
We are interested in orbits of slow-motion (compared with the speed 
of light) in the region where
\begin{equation}
\label{eq:R-region}
2m_0 =: R_S \ll R \ll R_H := c/H \,.
\end{equation}
The latter condition clearly covers all situations of practical 
applicability in the Solar System, since the Schwarzschild radius 
$R_S$ of the Sun is about 3~km $=2\cdot10^{-8}$AU and the 
`Hubble radius' $R_H$ is about $13.7\cdot 10^{9}$\,ly = $8.7\cdot 10^{14}$\,AU.

The approximation now consists in considering small perturbations of 
Keplerian orbits. Let $T$ be a typical timescale of the problem, like 
the period for closed orbits or else $R/v$ with $v$ a typical velocity. 
The expansion is then with respect to the following two parameters: 
\begin{subequations}
\label{eq:ExpParameters}
\begin{alignat}{3}
\label{eq:ExpParameters1}
&\varepsilon_{\sss 1} &&\,\approx\,\frac{v}{c}\,\approx\,
\left(\frac{m_0}{R}\right)^{\frac{1}{2}}
\quad &&\text{(slow-motion and weak-field)}\,,\\
\label{eq:ExpParameters2}
&\varepsilon_{\sss 2} &&\,\approx\,HT\quad 
&&\text{(small ratio of characteristic-time to world-age)}\,,
\end{alignat}

\end{subequations}
In order to make the expression to be approximated 
dimensionless we multiply (\ref{eq:TimeEquation}) by $1/c^2$ and 
(\ref{eq:McV-R-sec}) by $T^2/R$. Then we expand the right hand 
sides in powers of the parameters (\ref{eq:ExpParameters}), using the 
fact that $(HR/c)\approx\varepsilon_{\sss 1}\varepsilon_{\sss 2}$. 
From this and (\ref{eq:ConsAngMom}) we obtain (\ref{eq:impr-N-eqs}) 
if we keep only terms to zero-order in $\varepsilon_{\sss 1}$ and 
leading (i.e. quadratic) order in $\varepsilon_{\sss 2}$, where 
we also re-express $R$ as function of $t$. Note that in this 
approximation the areal radius $R$ is equal to the spatial geodesic
distance on the $t=\text{const.}$ hypersurfaces.


\section[Fully relativistic treatment for electromagnetically-bounded systems\\and the argument of Dicke and Peebles]{Fully~relativistic~treatment~for~electromagnetically-bounded systems and the argument of Dicke and Peebles}
\label{sec:DickePeebles}
In this section we show how to arrive at (\ref{eq:impr-N-eqs})
from a fully relativistic treatment of an electromagnetically 
bounded two-body problem in an expanding (spatially flat) universe.
This implies solving Maxwell's equations in the cosmological 
background~(\ref{eq:FlatFRW1}) for an electric point charge (the 
proton) and then integrate the Lorentz equations for the motion of
a particle (electron) in a bound orbit (cf. \cite{Bonnor:1999}). 
Equation (\ref{eq:impr-N-eqs}) then appears in an appropriate 
slow-motion limit. However, in oder to relate this straightforward 
method to a famous argument of Dicke \& Peebles, we shall proceed 
by taking a slight detour which makes use of the conformal properties 
of Maxwell's equations.

\subsection{The argument of Dicke and Peebles}
In reference~\cite{Dicke.Peebles:1964} Dicke \& Peebles presented 
an apparently very general and elegant argument, that purports
to show the insignificance of any dynamical effect of cosmological 
expansion on a local system that is either bound by electromagnetic or 
gravitational forces which should hold true \emph{at any scale}. 
Their argument involves a rescaling of spacetime 
coordinates, $(t,\vec x)\mapsto (\lambda t,\lambda\vec x)$ and 
certain assumptions on how other physical quantities, most prominently
mass, behave under such scaling transformations. For example, they 
assume mass to transform like $m\mapsto\lambda^{-1}m$. However, their 
argument is really independent of such assumptions, as we shall 
show below. We work from first principles to clearly display all 
assumptions made. 

We consider the motion of a charged point particle in an 
electromagnetic field. The whole system, i.e. particle plus 
electromagnetic field, is placed into a cosmological FRW-spacetime 
with flat ($k=0$) spatial geometry. The spacetime metric reads 
\begin{equation}
\label{eq:FlatFRW1}
g = c^2\, dt^2-a^2(t)\bigl(
dr^2+r^2\,(d\theta^2+\sin^2\theta\,d\varphi^2)
\bigr)\,.
\end{equation}
We introduce conformal time, $t_c$, via 
\begin{equation}
\label{eq:ConfTime} 
t_c=f(t):=\int^t_k\frac{dt'}{a(t')}\,,
\end{equation}
by means of which we can write (\ref{eq:FlatFRW1}) in a
conformally flat form, where $\eta$ denotes the flat Minkowski 
metric: 
\begin{equation}
\label{eq:FlatFRW2}
g=a_c^2(t_c)\bigl\{\underbrace{
c^2\,dt_c^2-dr^2-r^2\,(d\theta^2+\sin^2\theta\,d\varphi^2)}_%
{\displaystyle{\eta}}
\bigr\}\,.
\end{equation}
Here we wrote $a_c$ to indicate that we now expressed the expansion 
parameter $a$ as function of $t_c$ rather than $t$, i.e. 
\begin{equation}
\label{eq:A-c}
a_c:=a\circ f^{-1}\,.
\end{equation}
For example, if $a(t)=\sigma t^n$ ($0<n<1$), then we can choose $k=0$
in (\ref{eq:ConfTime}) and have 
\begin{equation}
\label{eq:TimeRelation1}
t_c=f(t)=\int_0^t\frac{dt'}{\sigma t'^n}=\frac{t^{1-n}}{\sigma(1-n)}\,,
\end{equation}
so that 
\begin{equation}
\label{eq:TimeRelation2}
t=f^{-1}(t_c)=\bigl[(1-n)\sigma\,t_c\bigr]^{1/(1-n)}\,,
\end{equation}
and therefore 
\begin{equation}
\label{eq:TimeRelation3}
a_c(t_c)=\alpha t_c^{n/(1-n)}\,,\quad\text{where}\quad
\alpha:=\bigl[(1-n)^n\sigma\bigr]^{1/(1-n)}\,.
\end{equation} 

The electromagnetic field is characterized by the tensor
$F_{\mu\nu}$, comprising electric and magnetic fields: 
\begin{equation}
\label{eq:EM-Components}
F_{\mu\nu}=
\begin{pmatrix}
0&E_n/c\\
-E_m/c&-\varepsilon_{mnj}B_j
\end{pmatrix}\,.
\end{equation}
In terms of the electromagnetic four-vector potential,
$A_\mu=(\varphi/c,-\vec A)$, one has 
\begin{equation}
\label{eq:FourPotential} 
F_{\mu\nu}=\partial_\mu A_\nu-\partial_\nu A_\mu
          =\nabla_\mu A_\nu-\nabla_\nu A_\mu\,,
\end{equation}
so that, as usual, $\vec E=-\vec\nabla\phi-\dot{\vec A}$. 
The expression for the four-vector of the Lorentz-force of a particle 
of charge $e$ moving in the field $F_{\mu\nu}$ is 
$e\,F^\mu_{\phantom{\mu}\alpha}u^\alpha$, where $u$ is the 
particle's four velocity. 

The equations of motion for the system Particle + EM-Field
follow from an action which is the sum of the action of the 
particle, the action for its interaction with the electromagnetic 
field, and the action for the free field, all placed in the 
background (\ref{eq:FlatFRW1}). Hence we write:  
\begin{equation}
\label{eq:ActionSum}
S=S_P+S_I+S_F\,,
\end{equation} 
where
\begin{subequations}
\begin{alignat}{2}
\label{eq:P-Action}
&S_P &&\,=\,-mc^2\int_z d\tau
       \,=\,-mc\int\sqrt{g(z',z')}\,d\lambda\,,\\
&S_I &&\,=\,-\,e\int_zA_\mu\,dx^\mu
       \,=\,-\,e\int A_\mu(z(\lambda))z'^\mu\,d\lambda\nonumber\\
\label{eq:I-Action}
&    &&\,=\,-\,\int d^4x\,A_\mu(x)\int\,d\lambda\ e\ 
            \delta^{(4)}(x-z(\lambda))\ z'^\mu\,,\\
\label{eq:F-Action}
&S_F &&\,=\,\frac{-1}{4}\int d^4x\,\sqrt{-\det g}\;
          g^{\mu\alpha}g^{\nu\beta}\,F_{\mu\nu}F_{\alpha\beta}
       \,=\,\frac{-1}{4}\int d^4x\,\eta^{\mu\alpha}
       \eta^{\nu\beta}\,F_{\mu\nu}F_{\alpha\beta}\,.
\end{alignat}
\end{subequations}
Here $\lambda$ is an arbitrary parameter along the worldline 
$z:\lambda\mapsto z(\lambda)$ of the particle, and $z'$
the derivative $dz/d\lambda$. The differential of the 
eigentime along this worldline is 
\begin{equation}
\label{eq:DefEigentime}
d\tau=\sqrt{g(z',z')}\,d\lambda
=\sqrt{g_{\mu\nu}(z(\lambda))\tfrac{dz^\mu}{d\lambda}
\tfrac{dz^\nu}{d\lambda}}\,d\lambda\,.
\end{equation}
It is now important to note that 1)~the background metric $g$
does not enter (\ref{eq:I-Action}) and that (\ref{eq:F-Action})
is conformally invariant (in 4 spacetime dimensions only!). Hence 
the expansion factor, $a(t_c)$, does not enter these two 
expressions. For this reason we could write (\ref{eq:F-Action})
in terms of the flat Minkowski metric, though it should be kept 
in mind that the time coordinate is now given by conformal time
$t_c$. This is \emph{not} the time read by standard clocks that 
move with the cosmological observers, which rather show the 
cosmological time $t$ (which is the proper time along the 
geodesic flow of the observer field $X=\partial/\partial t$). 

The situation is rather different for the action (\ref{eq:P-Action})
of the particle.  Its variational derivative with respect to
$z(\lambda)$ is 
\begin{equation}
\label{eq:VarDerS_p1}
\frac{\delta S_p}{\delta z^\mu(\lambda)}= -mc\,\left\{
\frac{\tfrac{1}{2}g_{\alpha\beta,\mu}\,z'^\alpha z'^\beta}{\sqrt{g(z',z')}} 
-\frac{d}{d\lambda}\left[\frac{g_{\mu\alpha}z'^\alpha}{\sqrt{g(z',z')}}
 \right]\right\} \, .
\end{equation} 
We now introduce the \emph{conformal proper time}, $\tau_c$, via 
\begin{equation}
\label{eq:CobfPropTime}
d\tau_c=(1/c)\,\sqrt{\eta(z',z')}\,d\lambda
       = (1/ca)\,\sqrt{g(z',z')}\,d\lambda\,.
\end{equation}
We denote differentiation with respect to $\tau_c$ by an overdot, 
so that e.g. $z'/\sqrt{g(z',z')}=\dot z/ca$. Using this to replace 
$z'$ by $\dot z\sqrt{g(z',z')}/ca$ and also $g$ by $a^2\eta$ in 
(\ref{eq:VarDerS_p1}) gives  
\begin{equation}
\label{eq:VarDerS_p2}
\frac{\delta S_p}{\delta z^\mu(\lambda)}= 
\frac{\sqrt{g(z',z')}}{ac}\,ma\,\left\{
\eta_{\mu\alpha}\ddot z^\alpha+P^\alpha_\mu\phi_{,\alpha}\right\}
\end{equation} 
where we set 
\begin{equation}
\label{eq:Abbrev}
a=:\exp(\phi/c^2)\quad\text{and}\quad
P^\alpha_\mu:=-\delta^\alpha_\mu+
\frac{\dot z^\alpha\dot z^\nu}{c^2}\eta_{\nu\mu}\,.
\end{equation}
Recalling that 
$\delta S_P=\int\frac{\delta S_p}{\delta z^\mu(\lambda)}\delta z^\mu d\lambda=
\int\frac{\delta S_p}{\delta z^\mu(\tau_c)}\delta z^\mu d\tau_c$ and 
using (\ref{eq:CobfPropTime}), (\ref{eq:VarDerS_p2}) is equivalent to 
\begin{equation}
\label{eq:VarDerS_p3}
\frac{\delta S_p}{\delta z^\mu(\tau_c)}= 
ma\bigl(\ddot z^\alpha+P^\alpha_\mu\phi_{,\alpha}\bigr)\,,
\end{equation} 
where from now on we agree to raise and lower indices using the 
Minkowski metric, i.e. $\eta_{\mu\nu}=\text{diag}(1,-1,-1,-1)$ in Minkowski 
inertial coordinates.
 
Writing (\ref{eq:I-Action}) in terms of the conformal proper time 
and taking the variational derivative with respect to $z(\tau_c)$ 
leads to $\delta S_I/\delta z^\mu(\tau_c)=-eF_{\mu\alpha}{\dot z}^\alpha$, 
so that 
\begin{equation}
\label{eq:VarDerSwrtZ}
\frac{\delta S}{\delta z^\mu(\tau_c)}
=ma\bigl(\ddot z_\mu+P^\alpha_\mu\phi_{,\alpha}\bigr)
-e\,F_{\mu\alpha}{\dot z}^\alpha\,.
\end{equation}

The variational derivative of the action with respect to the 
vector potential $A$ is 
\begin{equation}
\label{eq:VarDerSwrtA}
\frac{\delta S}{\delta A_\mu(x)}
=\partial_{\alpha}F^{\mu\alpha}(x)
-e\,\int d\tau_c\,\delta(x-z(\tau_c))\,\dot z^\mu(\tau_c)\,. 
\end{equation}

Equations (\ref{eq:VarDerSwrtZ}) and (\ref{eq:VarDerSwrtA}) 
show that the fully dynamical problem can be treated as if it were 
situated in static flat space. The field equations that follow 
from (\ref{eq:VarDerSwrtA}) are just the same as in Minkowski 
space. Hence we can calculate the Coulomb field as usual. On the 
other hand, the equations of motion receive two changes from the 
cosmological expansion term: the first is that the mass $m$ is 
now multiplied with the (time-dependent!) scale factor $a$, the 
second is an additional scalar force induced by $a$. 
Note that all spacetime dependent functions on the right hand 
side are to be evaluated at the particle's location $z(\tau_c)$, whose 
fourth component corresponds to $ct_c$. Hence, writing out all 
arguments and taking into account that the time coordinate is $t_c$, 
we have for the equation of motion
\begin{subequations}
\label{eq:DP-Motion}
\begin{alignat}{1}
\label{eq:DP-Motion1}
\ddot z^\mu
&\,=\,\frac{e}{ma_c(z^0/c)}\ F^\mu_{\phantom{\mu}\alpha}(z)\dot z^\alpha
- \bigl(-c^2\eta^{\mu\alpha}+\dot z^\mu\dot z^\alpha\bigr)
\partial_\alpha\ln a_c(z^0/c)\\
\label{eq:DP-Motion2}
&\,=\,\frac{e}{ma_c(z^0/c)}\ F^\mu_{\phantom{\mu}\alpha}(z)\dot z^\alpha
- \bigl(-c\eta^{\mu 0}+\dot z^\mu\dot z^0/c\bigr)\,a'_c(z^0/c)/a_c(z^0/c)\,,  
\end{alignat}    
\end{subequations}
where $a'_c$ is the derivative of $a_c$.

So far no approximations were made. Now we write $\dot z^\mu=\gamma(c,\vec v)$,
where $\vec v$ is the derivative of $\vec z$ with respect to the 
conformal time $t_c$, henceforth denoted by a prime, and 
$\gamma=1/\sqrt{1-v^2/c^2}$. Then we specialize to slow motions, i.e. 
neglect effects of quadratic or higher powers in $v/c$ (special
relativistic effects). For the spatial part of (\ref{eq:DP-Motion2}) we get
\begin{equation}
\label{eq:DP-Motion3}
\vecpp{z} + \vecp{z}\ (a'_c/a_c)
=\frac{e}{ma_c}\bigl( \vec E + \vecp{z} \times \vec B \bigr)\,,
\end{equation}
where we once more recall that the spatial coordinates used here are 
the comoving (i.e. conformal) ones and the electric and 
magnetic fields are evaluated at the particle's position $\vec z(t_c)$. 

From the above equation we see that the effect of cosmological expansion
in the conformal coordinates shows up in two ways: first in a time 
dependence of the mass which scales with $a_c$, and, second, in the
presence a friction term. Let us, for the moment, neglect the 
friction term. In the adiabatic approximation, which is justified 
if typical time scales of the problem at hand are short compared to 
the world-age (corresponding to small $\varepsilon_{\sss 2}$ in 
(\ref{eq:ExpParameters2})), the time-dependent mass term leads to 
a time varying radius in comoving (or conformal) coordinates of 
$r(t_c)\propto 1/a_c(t_c)$. Hence the physical radius (given by the 
cosmological geodesically spatial distance), $r_*=a_c r$, stays 
constant in this approximation. In this way Dicke \& Peebles 
concluded in \cite{Dicke.Peebles:1964} that electromagnetically 
bound systems do not feel \emph{any} effect of cosmological expansion.

Let us now look at the effect of the friction term which the 
analysis of Dicke \& Peebles neglects. It corresponds to the 
decelerating force $-\vec v a'_c/a_c$ which e.g. for the simple 
power-law expansion (\ref{eq:TimeRelation3}) becomes 
\begin{equation}
\label{eq:FritionTerm}
-\vec v \tfrac{n}{n-1}t^{-1}_c
=-\vec v\,n\sigma\, t^{n-1}\,.
\end{equation} 
Clearly it must cause any stationary orbit to decay. For example, 
as a standard first-order perturbation calculation shows, a circular 
orbit of radius $r$ and angular frequency (with respect to 
conformal time $t_c$) $\omega_c$ will suffer a relative decay per 
revolution of
\begin{equation}
\label{eq:RadialDecay} 
\frac{\Delta r}{r}\Big\vert_{\rm revol.}=-\,\frac{a'_c/a_c}{3\omega_c}\,.
\end{equation} 
Recall that this is an equation in the (fictitious) Minkowski space
obtained after rescaling the physical metric. However, it equates 
two scale invariant quantities. Indeed, the relative length change 
$\Delta r/r$ is certainly scale invariant and so is the 
relative length change per revolution. On the right hand side we 
take the quotient of two quantities which scale like an inverse time. 
In physical spacetime, coordinatized by cosmological time $t$, the 
right hand side becomes $-\,H/3\omega$, where as usual $H=\dot a/a$ 
and $\omega$ is the angular frequency with respect to $t$. Hence the 
relative radial decay in physical space is of the order of the ratio
between the orbital period and the inverse Hubble constant 
(`world-age'), i.e. of order $\varepsilon_{\sss 2}$ 
(cf.(\ref{eq:ExpParameters2})).

Since the friction term contributes to the leading-order effect of 
cosmological expansion, we conclude that the argument of Dicke \& 
Peebles, which neglects this term, is not sufficient to estimate 
such effects.

\subsection{Equations of motion in the physical coordinates}

We now show that (\ref{eq:impr-N-eqs}) is indeed arrived at if the 
friction term is consistently taken into account. To see this we merely
need to rewrite equation~(\ref{eq:DP-Motion3}) in terms of the physical 
coordinates given by the cosmological time $t$ and the cosmological 
geodesic spatial distance $r_*:=a(t)r$. We have $dt_c/dt=1/a$ 
and the spatial geodesic coordinates are $\vec y:= a(t) \vec z$. 
Denoting by an overdot the time derivative with respect to $t$, 
the left hand side of~(\ref{eq:DP-Motion3}) becomes
\begin{equation}
\label{eq:backTransformation}
\vecpp{z} + \vecp{z}\ (a'_c/a_c) = a\ \ddot{{\vec y}} - \ddot a\ \vec y\,.
\end{equation}
This shows that the friction term in the unphysical coordinates
becomes, in the physical coordinates, the familiar acceleration 
term~(\ref{eq:cosmological-acc}) due to the Hubble-law.
Dividing by $a$ equation~(\ref{eq:DP-Motion3}) and inserting 
$\vec E(\vec{z}) = Q\vec{z}/|\vec{z}\,|^3$ and $\vec B(\vec{z}) = 0$, we get
\begin{equation}
\label{eq:DP-Motion4}
\ddot{{\vec y}} - \vec{y}\ (\ddot a/a) = \frac{eQ}{m |\vec{y}|^3}\vec{y} \,.
\end{equation}
Finally, introducing polar coordinates in the orbital plane
we exactly get~(\ref{eq:impr-N-eqs}).


\section{Kinematical effects}
\label{sec:Rosales}
It has been suggested in \cite{Rosales.Sanchez-Gomez:1998} 
and again in \cite{Rosales:2002} that there may be significant 
\emph{kinematical} effects that may cause apparent anomalous 
acceleration of spacecraft orbits in an expanding cosmological 
environment. More precisely it was stated that there is an additional 
acceleration of magnitude $Hc\approx 0.7\cdot 10^{-9}\textrm{m}/\textrm{s}^2$,
which is comparable to the measured anomalous acceleration of the 
Pioneer spacecrafts. The cause of such an effect lies in the way one 
actually measures spatial distances and determines the clock readings 
they are functions of (a trajectory is a `distance' for each given 
`time'). The point is this: equations of motions give us, for example, 
simultaneous (with respect to cosmological time) spatial geodesic 
distances as functions of cosmological time. This is what we implicitly 
did in the Newtonian analysis. But, in fact, spacecraft ranging is done 
by exchanging electromagnetic signals. The notion of spatial distance 
as well as the notion of simultaneity introduced thereby is \emph{not}
the same. Hence the analytical expression of the `trajectory' so 
measured will be different. 
      
To us this seems an important point and the authors of 
\cite{Rosales.Sanchez-Gomez:1998} and \cite{Rosales:2002} were well 
justified to draw proper attention to it. However, we will now explain 
why we do not arrive at their conclusion. Again we take care to state 
all assumptions made. 
\subsection{Local Einstein-simultaneity in general spacetimes}
We consider a general Lorentzian manifold $(\M,g)$ as spacetime. 
Our signature convention is $(+,-,-,-)$ and $ds$ is taken to 
have the unit of length. The differential of eigentime is $d\tau=ds/c$.  
In general coordinates $\{x^\mu\}$, the metric reads 
\begin{equation}
\label{eq:MetricGen}
ds^2=g_{\mu\nu}dx^\mu\,dx^\nu=g_{tt}dt^2+2g_{ta}dt\,dx^a+g_{ab}dx^a\,dx^b\,.
\end{equation}
The observer at fixed spatial coordinates is given by the vector 
field (normalized to $g(X,X)=c^2$)
\begin{equation}  
\label{eq:observer}
X=\frac{c}{\sqrt{g_{tt}}}\frac{\partial}{\partial t}\,.
\end{equation}
Consider the light cone with vertex $p\in\M$; one has $ds^2=0$, which 
allows to solve for $dt$ in terms of the $dx^a$ (all functions $g_{ab}$
are evaluated at $p$, unless noted otherwise): 
\begin{equation}
\label{eq:LightConeDiff}
dt_{1{,}2}=-\,\frac{g_{ta}}{g_{tt}}\,dx^a\pm\sqrt{\left(
\frac{g_{ta}g_{tb}}{g^2_{tt}}-\frac{g_{ab}}{g_{tt}}
\right)dx^a\,dx^b}\,.
\end{equation}
The plus sign corresponds to the future light-cone at $p$, 
the negative sign to the past light cone. An integral line of $X$ 
in a neighbourhood of $p$ cuts the light cone in two points,
$q_+$ and $q_-$. If $t_p$ is the time assigned to $p$, then 
$t_{q_+}=t_p+dt_1$ and $t_{q_-}=t_p+dt_2$. The coordinate-time 
separation between these two cuts is $t_{q_+}-t_{q_-}=dt_1-dt_2$,
corresponding to a proper time $\sqrt{g_{tt}}(dt_1-dt_2)/c$ for 
the observer $X$. This observer will associate 
a \emph{radar-distance} $dl_*$ to the event $p$ of $c/2$ times that 
proper time interval, that is: 
\begin{equation}
\label{eq:RadarDist}
dl_*^2=h=
\left(\frac{g_{ta}g_{tb}}{g_{tt}}-g_{ab}\right)dx^a\,dx^b\,.
\end{equation}

The event on the integral line of $X$ that the observer will 
call Einstein-synchronous with $p$ lies in the middle between 
$q_+$ and $q_-$. Its time coordinate is in first-order approximation 
given by $\tfrac{1}{2}(t_{q_+}+t_{q_-})=t_p+\tfrac{1}{2}(dt_1+dt_2)=t_p+dt$,
where  
\begin{equation}
\label{eq:EinstSynShift}
dt:=\tfrac{1}{2}(dt_1+dt_2)=-\frac{g_{ta}}{g_{tt}}\,dx^a\,.
\end{equation}

This means the following: the Integral lines of $X$ are parameterized 
by the spatial coordinates $\{x^a\}_{a=1,2,3}$. Given a point $p$, 
specified by the orbit-coordinates $x^a_p$ and the time-coordinate 
$t_p$, we consider a neighbouring orbit of 
$X$ with orbit-coordinates $x^a_p+dx^a$. The event on the latter which 
is Einstein synchronous with $p$ has a time coordinate $t_p+dt$, where 
$dt$ is given by (\ref{eq:EinstSynShift}), or equivalently 
\begin{equation}
\label{eqSynchConnForm}
\theta:=dt+\frac{g_{ta}}{g_{tt}}\,dx^a=0\,.
\end{equation}
Using a differential geometric language we may say that
Einstein simultaneity defines a \emph{distribution} $\theta=0$. 

The metric (\ref{eq:MetricGen}) can be written in terms of the 
radar-distance metric $h$ (\ref{eq:RadarDist}) and the simultaneity 
1-form $\theta$ as follows: 
\begin{equation}
\label{eq:MetricGenAlt}
ds^2=g_{\mu\nu}dx^\mu\,dx^\nu=g_{tt}\,\theta^2-h\,,
\end{equation}
showing that the radar-distance is just the same as the 
Einstein-simultaneous distance. A curve $\gamma$ in $\M$ intersects 
the flow lines of $X$ perpendicularly iff $\theta(\dot\gamma)=0$, 
which is just the condition that neighbouring clocks along 
$\gamma$ are Einstein synchronized.

\subsection{Application to isotropic cosmological metrics}
We consider isotropic cosmological metrics. In what follows we drop 
for simplicity the angular dimensions. Hence we consider metrics of the 
form
\begin{equation}
\label{eq:CosMetric}
ds^2=c^2dt^2-a(t)^2dr^2\,.
\end{equation} 
The expanding observer field is 
\begin{equation}
\label{eq:GeodObs}
X=\frac{\partial}{\partial t}\,.
\end{equation} 
The Lagrangian for radial geodesic motion is 
$L=\tfrac{1}{2}\bigl(c^2{\dot t}^2-a^2{\dot r}^2\bigr)$, 
leading to the only non-vanishing Christoffel symbols
\begin{equation}
\label{eq:ChristSymb}
\Gamma_{rr}^t=\frac{a\dot a}{c^2}\,,\quad
\Gamma_{tr}^r=\frac{\dot a}{a}=:H\,.
\end{equation}
Hence $X$ is geodesic, since 
\begin{equation}
\nabla_XX=\Gamma_{tt}^\mu\partial_\mu=0\,.
\end{equation}

On a hypersurface of constant $t$ the radial geodesic distance is 
given by $ra(t)$. Making this distance into a spatial coordinate,
$r_*$, we consider the coordinate transformation 
\begin{equation}
\label{eq:CoordTrans}
t\mapsto t_*:=t\,,\quad
r\mapsto r_*:=a(t)r\,.
\end{equation}
The field $\partial/\partial t_*$ is given by 
\begin{equation}
\label{eq:TstarVF}
\frac{\partial}{\partial t_*}
=\frac{\partial}{\partial t}-Hr\,\frac{\partial}{\partial r}.
\end{equation}
In contrast to (\ref{eq:GeodObs}), whose flow connects co-moving 
points of constant coordinate $r$, the flow of (\ref{eq:TstarVF}) 
connects points of constant geodesic distances, as measured in 
the surfaces of constant cosmological time. This could be called 
\emph{cosmologically instantaneous geodesic distance}. It is now very important 
to realize that this notion of distance is not the same as the radar 
distance that one determines by exchanging light signals in the usual 
(Einsteinian) way. Let us explain this in detail: 

From (\ref{eq:CoordTrans}) we have $adr=dr_*-r_*Hdt$, where 
$H:=\dot a/a$ (Hubble parameter). Rewriting the metric 
(\ref{eq:CosMetric}) in terms of $t_*$ and $r_*$ yields
\begin{alignat}{2}
& ds^2&&\,=\,
 c^2(1-(Hr_*/c)^2)\,dt_*^2-dr_*^2+2Hr_*\,dt\,dr_*\nonumber\\
\label{eq:MetricNewCoord}
&     &&\,=\,
\underbrace{c^2\Bigl\{1-(Hr_*/c)^2\Bigr\}}_{\textstyle g_{t_*t_*}}
 \Bigl\{\underbrace{dt_*+\frac{Hr_*/c^2}{1-(Hr_*/c)^2}\,dr_*}_{%
\textstyle\theta}\Bigr\}^2
-\underbrace{\frac{dr_*^2}{1-(Hr_*/c)^2}}_{\textstyle h}\,,
\end{alignat}
Hence the differentials of radar-distance and  time-lapse for 
Einstein-simultaneity are given by  
\begin{subequations}
\label{eq:DiffStarQuant}
\begin{alignat}{2}
\label{eq:DiffRadDist}
& dl_*&&\,=\,\frac{dr_*}{\sqrt{1-(Hr_*/c)^2}}\,,\\
\label{eq:DiffLapseEinstSim}
& dt_*&&\,=\,-\,\frac{Hr_*/c^2}{1-(Hr_*/c)^2}dr_*\,.
\end{alignat}
\end{subequations}

Let the distinguished observer (us on earth) now move along the 
geodesic $r_*=0$. Integration of (\ref{eq:DiffStarQuant}) from $r_*=0$
to some value $r_*$ then gives the radar distance $l_*$ as well as 
the time lapse $\Delta t_*$ as functions of the cosmologically 
simultaneous geodesic distance $r_*$: 
\begin{subequations}
\label{eq:IntStarQuant}
\begin{alignat}{3}
\label{eq:IntRadDist}
& l_*&&\,=\,(c/H)\ 
  \sin^{-1}(H\,r_*/c)
  &&\,\approx\,r_*\bigl\{1+\tfrac{1}{6}(Hr_*/c)^2 +O(3)\bigr\}\\
\label{eq:IntLapseEinstSim1}
& \Delta t_*&&\,=\,(1/2H)\, 
  \ln\bigl(1-(H\,r_*/c)^2\bigr)
  &&\,\approx\,(r_*/c)\bigl\{-\tfrac{1}{2}(Hr_*/c)+O(2)\bigr\}
\end{alignat}
\end{subequations}
Combining both equations in (\ref{eq:IntStarQuant}) allows to 
express the time-lapse in terms of the radar-distance: 
\begin{equation}
\label{eq:IntLapseEinstSim2}
\Delta t_*=H^{-1}\,\ln\bigl(\cos(H\,l_*/c)\bigr)
\approx(l_*/c)\bigl\{-\tfrac{1}{2}(Hl_*/c)+O(2)\bigr\}\,.
\end{equation}

\begin{figure}
\centering\epsfig{figure=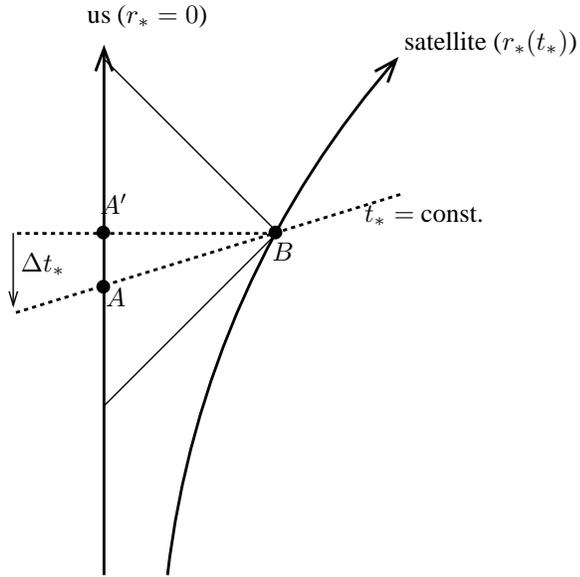,width=0.4\linewidth}
\put(-113,103){\small$A$}
\put(-115,138){\small$A'$}
\put(-50,120){\small$B$}
\put(-120,210){\small us\ ($r_*=0$)}
\put(-0,200){\small satellite\ ($r_*(t_*)$)}
\put(-15,135){\small $t_*=$ const.}
\put(-145,116){\small $\Delta t_*$}
\caption{\label{fig:SatOrbit}
The observer (`us') moves on the geodesic worldline $r_*=0$ and measures 
the spatial distance to a satellite by exchanging electromagnetic signals
(radar coordinates). 
The satellite's worldline is analytically described by a function 
$r_*(t_*)$. The surface of constant cosmological time $t_*$ (tilted 
dashed line) intersects our worldline at $A$ and the satellites 
worldline at $B$. The coordinate $r_*$ corresponds to the geodesic 
distance in that hypersurface of constant cosmological time, i.e. to 
the length of $\overline{AB}$ as measured in the spacetime geometry. 
However, using radar coordinates, the observer defines $B$ to be 
simultaneous to his event $A'$ and attributes to it the distance 
$l_*$. Hence instead of $r_*(t_*)$ he uses $l_*(r_*(t_*-\Delta t_*))$, 
where $l_*(r_*)$ is given by (\ref{eq:IntRadDist}). This leads to 
(\ref{eq:ObserverCurve}).}
\end{figure}

Now, suppose a satellite $S$ moves on a worldline $r_*(t_*)$ 
in the neighbourhood of our worldline $r_*=0$. Assume that we measure 
the distance to the satellite by radar coordinates. Then instead of 
the value $r_*$ we would use $l_*$ and instead of the argument $t_*$ 
we would assign the time $t_*-\Delta t_*$ which corresponds to the 
value of cosmological time at that event on our worldline that is 
Einstein synchronous to the event $(t_*,r_*)$; see Figure\,\ref{fig:SatOrbit}.

Hence we have 
\begin{subequations}
\label{eq:ObserverCurve}
\begin{alignat}{1}
\label{eq:ObserverCurve1}
l_*(t_*)&\,=\,(c/H)\ \sin^{-1}\bigl\{r_*(t_*+\Delta t_*)H/c\bigr\}\\
\label{eq:ObserverCurve2}
&\,\approx\, r_*-\tfrac{1}{2}(v/c)(Hc)(r_*/c)^ 2\,,
\end{alignat}
\end{subequations}
where (\ref{eq:ObserverCurve2}) is (\ref{eq:ObserverCurve1}) to leading 
order and all quantities are evaluated at $t_*$. We set $v=\dot r_*$. 

To see what this entails we Taylor expand in $t_*$ around $t_*=0$ 
(just a convenient choice): 
\begin{equation}
\label{eq:TaylorExpOrbit1} 
r_*(t_*)=r_0+v_0t_*+\tfrac{1}{2}a_0t_*^ 2 +\cdots
\end{equation}
and insert in (\ref{eq:ObserverCurve2}). This leads to 
\begin{equation}
\label{eq:TaylorExpOrbit2} 
l_*(t_*)=\tilde r_0+\tilde v_0t_*+\tfrac{1}{2}\tilde a_0t_*^ 2 +\cdots\,,
\end{equation}
where,  
\begin{alignat}{2}
\label{eq:TaylorExpOrbit3a}
& \tilde r_0&&\,=\,
r_0-(Hc)\ \tfrac{1}{2}(v_0/c)(r_0/c)^2\\
\label{eq:TaylorExpOrbit3b}
& \tilde v_0&&\,=\,
v_0-(Hc)\ (v_0/c)^2(r_0/c)\\
\label{eq:TaylorExpOrbit3c}
& \tilde a_0&&\,=\,
a_0-(Hc)\ \bigl\{(v_0/c)^3+(r_0/c)(v_0/c)(a_0/c)\bigl\}
\end{alignat}
These are, in quadratic approximation, the sought-after relations 
between the quantities measured via radar tracking (tilded) and the 
quantities which arise in the (improved) Newtonian equations of motion 
(not tilded). 

The last equation (\ref{eq:TaylorExpOrbit3c}) shows that there is an 
apparent inward pointing acceleration, given by $Hc$ times the 
$(v/c)^3+\cdots$ term in curly brackets. The value of $Hc$ is indeed of 
the same order of magnitude as the anomalous Pioneer acceleration, as 
emphasized in \cite{Rosales.Sanchez-Gomez:1998,Rosales:2002}. 
However, in contrast to these authors, we do get the additional term 
in curly brackets, which in case of the Pioneer spacecraft suppresses 
the $Hc$ term by 13 orders of magnitude! Hence, according to our 
analysis, and in contrast to what is stated in 
\cite{Rosales.Sanchez-Gomez:1998,Rosales:2002}, there is 
no significant kinematical effect resulting from the distinct 
simultaneity structures inherent in radar and cosmological 
coordinates. We should stress, however, that this verdict is strictly 
limited to our interpretation of what the kinematical effect actually 
consists in, which is most concisely expressed in 
(\ref{eq:ObserverCurve}).\footnote{Our equation 
(\ref{eq:IntLapseEinstSim2}) corresponds to equation~(10) of 
\cite{Rosales.Sanchez-Gomez:1998}. From it the authors of 
\cite{Rosales.Sanchez-Gomez:1998} and \cite{Rosales:2002} 
immediately jump to the conclusion that there is ``an effective 
residual acceleration directed toward the centre of coordinates; 
its constant value is $Hc$''. We were unable to understand how this 
conclusion is reached. Our interpretation of the meaning of 
(\ref{eq:ObserverCurve}) does not support this conclusion.}


\section{Summary and outlook}
\label{sec:Summary}
We think it is fair to say that there are no theoretical hints that 
point towards a \emph{dynamical} influence of cosmological expansion 
comparable in size to that of the anomalous acceleration of the 
Pioneer spacecrafts. There seems to be no controversy over this 
point, though for completeness it should be mentioned that according 
to a recent suggestion \cite{Palle:2005} it might become relevant for 
future missions like LATOR. This suggestion is based on the model of 
Gautreau~\cite{Gautreau:1984b} which, as already mentioned in 
Section~\ref{sec:McVittie}, we find hard to relate to the problem 
discussed here. Rather, as the $(\ddot a/a)$--improved Newtonian 
analysis in Section~\ref{sec:NewtonianApproach} together with its 
justification given in Sections~\ref{sec:McVittie} and 
\ref{sec:DickePeebles} strongly suggests, there is no genuine 
relativistic effect coming from cosmological expansion at the 
levels of precision envisaged here.   

On the other hand, as regards \emph{kinematical} effects, the situation 
is less unanimous. It is very important to unambiguously understand 
what is meant by `mapping out a trajectory', i.e. how to assign `times' 
and `distances'. Eventually we compare a functional relation between 
`distance' and `time' with observed data. That relation is obtained by 
solving some equations of motion and it has to be carefully checked 
whether the methods by which the tracking data are obtained match the 
interpretation of the coordinates in which the analytical problem is 
solved. In our way of speaking dynamical effects really influence the 
worldline of the object in question, whereas kinematical effects 
change the way in which one and the same worldline is mapped out 
from another worldline representing the observer. 

The latter problem especially presents itself in a time dependent 
geometry of spacetime. Mapping out a trajectory then becomes dependent 
on ones definition of `simultaneity' and `simultaneous spatial distance'
which cease to be unique. An intriguing suggestion has been made 
\cite{Rosales.Sanchez-Gomez:1998,Rosales:2002} that the PA is merely 
a result of such an ambiguity. However, our analysis suggests that no 
significant relativistic effects result within the Solar System, 
over and above those already taken into account, as e.g. the Shapiro 
time delay. 

What has been said so far supports the view that there is no interesting 
impact of cosmological expansion on the specific problem of 
\emph{satellite navigation} in the Solar System. However, turning now to a 
more general perspective, the problem of how local inhomogeneities on
a larger scale affect, and are affected by, cosmological expansion is
of utmost importance. Many scientific predictions concerning cosmological 
data rely on computations within the framework of the standard 
homogeneous and isotropic models, without properly estimating the 
possible effects of local inhomogeneities. Such an estimation would 
ideally be based on an exact inhomogeneous solution to Einstein's 
equations, or at least a fully controlled approximation to such a 
solution.  The dynamical and kinematical impact of local 
inhomogeneities might essentially influence our interpretation of 
cosmological observations. As an example we mention recent serious 
efforts to interpret the same data that are usually taken to prove the 
existence of a positive cosmological constant $\Lambda$ in a context 
with realistic inhomogeneities where $\Lambda=0$; see~\cite{Celerier:2000} 
and \cite{Mansouri:2005}. 

To indicate possible directions of research, we stress again that 
there are several approaches to the problem of how to rigorously 
combine an idealized local inhomogeneity---a single star in the 
most simple case---with an 
homogeneous and isotropic cosmological background. We mentioned 
that of Einstein \& Straus~\cite{Einstein.Straus:1945} and its 
refinement by Sch\"ucking~\cite{Schuecking:1954}, that of  
Gautreau~\cite{Gautreau:1984b}, that by 
Bonnor~\cite{Bonnor:1996,Bonnor:1999,Bonnor:2000a,Bonnor:2000b}, 
and especially the classic work by McVittie~\cite{McVittie:1933}
that was later elaborated on by Hogan~\cite{Hogan:1990} and properly 
interpreted by the penetrating analysis of 
Nolan~\cite{Nolan:1998,Nolan:1999a,Nolan:1999b}. Nolan also showed 
that only in the case $k=0$ does McVittie's solution 
represent a central mass embedded into a cosmological background. 
The problem is that, due to the non-linearity of Einstein's 
equations, the spherical inhomogeneity does not show just as an 
addition to the background. Hence a notion of quasi-local mass 
has to be employed in order to theoretically detect local mass 
abundance. However, as is well known, it is a notoriously difficult 
problem in General Relativity to define a physically appropriate 
notion of quasi-local mass. Workable definitions only exist in special 
circumstances, as for example in case of spherical symmetry, where 
the concept of Misner-Sharp mass can be employed, as explained in 
Section~\ref{sec:McVittieInterpretation}.

As a project for future research we therefore suggest to further 
probe and develop applications of the spatially flat ($k=0$) McVittie 
solution, taking due account of recent progress in our theoretical 
understanding of it. As a parallel development, the implications 
of Gautreau's model should be developed to an extent that allows 
their comparison with those of McVittie's model. 

\subsubsection*{Acknowledgements}

 This work was supported by the European Space Agency (ESA) under the Ariadna 
 scheme of the Advanced Concepts Team, contract 18913/05/NL/MV. 
 We are grateful to the ESA, Andreas Rathke, and Nicholas Lan for 
 support, discussions, and patience. We also thank Claus L\"ammerzahl, 
 Hartmann R\"omer, and Sebastian Schlicht for helpful discussions and 
 pointing out relevant references.


\newpage
\appendix
\section{Additional material}
\label{sec:AdditionalMaterial}
In this section we collect some background information which was 
implicitly used throughout the text.

\subsection{A simple estimate of the dynamical effect of cosmological expansion}
The radial acceleration due to the Newtonian Sun attraction is 
given by: 
\begin{equation}
\ddot{r}\,|_{\rm\sss Sun} = - \frac{G M_\odot}{r^2} \,
            \approx \,-\frac{60}{(r/10\,{\rm AU})^2} \,10^{-6}\,{\rm m/s^2}\,.
\end{equation}
The geodesic distance, $r_*$, between two freely falling bodies in an 
expanding FRW universe~(\ref{eq:FlatFRW1}), measured on a hypersurface 
of constant cosmological time, varies in time according to the Hubble-law 
$\dot{r}_*=H r_*$. The related `acceleration' is then 
$\ddot{r}_*=\dot{r}_*H+r_*\dot{H}=r_*(H^2+\dot{H})=r_*\,\ddot{a}/a=-qH^2r_*$.
At the present time (see Section~\ref{sec:astro-data}) we have
(suppressing the asterisk):
\begin{equation}
\label{eq:expansion-dynamical}
\ddot{r}\,|_{\rm\sss cosm. acc.} = \,\frac{\ddot a}{a}\,r \,\approx\,
 4\,(r/10\,{\rm AU})\,10^{-24}\,{\rm m/s^2}\,.
\end{equation}
This naive derivation of the \emph{dynamical} effect of the cosmological
expansion may be confirmed by the fully general relativistic
treatment, as showed in Section~\ref{sec:McVittie}. Notice that, according 
to our measurements, the universe is presently in a phase of accelerated 
expansion, hence~(\ref{eq:expansion-dynamical}) results in an acceleration 
pointing away from the Sun. This is in the \emph{opposite} direction of the 
Pioneer--effect~(\ref{eq:anomalous-acceleration}) and also smaller by 14 
orders of magnitude.

\subsection{The Sch\"ucking radius for different astronomical scales}
\label{sec:SchueckingAtScales}
In the following we evaluate the radius of the Sch\"ucking 
vacuole~(\ref{eq:Schuecking-radius}) for various characteristic 
central masses. Using as cosmological matter density 
$\rho_{\rm m}=\Omega_{\rm m}\rho_{\rm c}$ 
(see Appendix~\ref{sec:astro-data}), we get:
\begin{itemize}
\item {\bf Solar System scale:} 
$r_S(M_\odot) = 570$ ly, which is much larger than the average 
distance between stars in the Milky Way, being about 10 ly.
\item {\bf Galaxy scale:} 
$r_S(M_{\rm MW}) = 3\!-\!4$ Mly, which is again too big 
since this would also include other galaxies such as the Large and the Small 
Magellanic Cloud, as well as several dwarf galaxies.
\item {\bf Cluster scale:} 
$r_S(M_{\rm LG}) = 5\!-\!7$ Mly, which is just about the threshold 
since the nearest object not belonging to this cluster is NGC 55 
(a galaxy belonging to the Sculptor Group) about 5 Mly away.
\item {\bf Supercluster scale:} 
$r_S(M_{\rm VSC}) = 57$ Mly, which is inside the Virgo Supercluster 
whose radius is about 100 Mly.
\end{itemize}
This shows that the Einstein-Straus matching works at best from 
and above cluster scale. 

\subsection{Some astronomical and cosmological data}
\label{sec:astro-data}
For the convenience of the readers, we collect some relevant 
numerical information. 

{\bf Length units}\\
1 AU = $149.6 \cdot 10^6$ km = $1.5 \cdot 10^{11}$ m 
     = 492 ls = 8.2 lmin = $1.58 \cdot 10^{-5}$ ly \\
1 pc = 3.26 ly 

{\bf Time units}\\
1 yr = $3.1 \cdot 10^7$ s \\
Age of the Universe $\approx 13.7 \cdot 10^9$ yr = $4.32 \cdot 10^{17}$ s

{\bf Velocity units}\\
1 AU/yr = 4.74 km/s

{\bf Mass units}\\
1 $M_{\odot} = 2 \cdot 10^{30}$ kg 
(= sometimes referred to as twice the mass of the average star in 
the Milky Way)

\subsection*{The Universe}
{\bf Our galaxy: Milky Way (MW)}\\
Number of stars in the MW: $2\!-\!4\cdot 10^{11}$ \\
$M_{\mathrm{MW}}$ = $2\!-\!4\cdot 10^{11} M_{\odot}$ \\
Diameter: 100 kly \\
Average distance between stars in the MW: 10 ly \\
Nearest galaxy: Large Magellanic Cloud. 
Mass: $10^{10} M_{\odot}$, distance: 170 kly, diameter:~30~kly\\

{\bf Our cluster: Local Group (LG)}\\
Number of stars in the LG: $7 \cdot 10^{11}$ \\
$M_{\mathrm{LG}} = 7\!-\!20 \cdot 10^{11} M_{\odot}$ \\
Diameter: 10 Mly \\
Nearest clusters: Sculptor Group (distance: 10 Mly) and
Maffei 1 Groups (distance: 10 Mly). \\
Nearest galaxy: NGC55 (Sculptor Group), distance: 5 Mly.\\

{\bf Our supercluster: Virgo Supercluster (VSC)}\\
Number of stars in the VSC: $2 \cdot 10^{14}$ \\
$M_{\mathrm{VSC}} = 1 \cdot 10^{15} M_{\odot}$ \\
Diameter: 200 Mly \\
Nearest supercluster: Centaurus (distance: 200 Mly) \\
Nearest cluster: A3526 (Centaurus Supercluster), distance: 142 Mly \\

\subsection*{Cosmology data}
Hubble parameter (today): 
$H_0=h_0 \cdot 100$ km s$^{-1}$ Mpc$^{-1}$ = $h_0 /(3.08 \cdot 10^{17}$s),\\
where $h_0 = 0.7$

Critical density (today):
$\rho_c=3H_0^2/8 \pi G= h_0^2 \cdot 1.89 \cdot 10^{-29}$ g/cm$^3$ 

Definition of the cosmological parameters:
\begin{align}
   &\Omega_m:=\rho_m/\rho_c\,, \\
   &\Omega_\Lambda:=\rho_\Lambda/\rho_c=\Lambda/3 H_0^2\,, \\
   &\Omega_k:=-kc^2/(H_0 a_0)^2\,.
\end{align}
One has $\Omega_m + \Omega_\Lambda + \Omega_k = 1$ (cosmological triangle),
where todays values are given by: 
\begin{equation}
   (\Omega_m, \Omega_\Lambda, \Omega_k) \approx (1/3, 2/3, 0)\,.
\end{equation}
Thus from $q_0 = (1/2)\Omega_m - \Omega_\Lambda$ it follows that 
$q_0 \approx -1/2$.\\

\subsection*{Pioneer 10 and 11 data}
The following data are taken from~\cite{Anderson.etal:2002a}.
\setlength{\tabcolsep}{0.3cm}
\begin{table}[hbt]
\centering
\begin{tabular}{|l|c|c|}
\hline
\null                 &       P10      &      P11      \\
\hline 
Launch                & 2 Mar 1972     &  5 Apr 1973   \\
\hline
Planetary             & Jupiter: 4 Dec 1973  & Jupiter: 2 Dec 1974  \\
encounters            &       \null          & Saturn:  1 Sep 1979  \\
\hline
Tracking data         & \,\,3 Jan 1987 -- 22 Jul 1998 & 5 Jan 1987 -- 1 Oct 1990 \\
Distance from the Sun & 40 AU -- 70 AU       & 22.4 AU -- 31.7 AU   \\
Light round-trip time &  11 h -- 19 h        & 6 h     -- 9 h       \\
Radial velocity       & 13.1 Km/s -- 12.6 Km/s & N.A.               \\
\hline
\end{tabular}
\caption{Some orbital data of the Pioneer 10 and 11 spacecrafts.}
\label{pioneer-data}
\end{table}

{\bf Tracking system}\\
Uplink frequency as received from Pioneer (approx.): $\nu_{u,2}=2.11$ GHz.\\
Downlink frequency emitted from Pioneer: $\nu_{d,2}=T\nu_{u,2}=2.292$ GHz.\\
Spacecraft transponder turnaround ratio: $T=240/221$.\\

{\bf Measured effect}\\ 
Measured is an almost constant residual (meaning after subtraction of 
all the known effects) frequency drift of the received tracking signal.
The drift is a blue-shift at the constant rate 
\begin{equation}
  \label{eq:anomalous-blue-shift}
  \dot{\nu} = (5.99 \pm 0.01)\,10^{-9}\,{\rm Hz/s}
\end{equation}
which, if interpreted as a special-relativistic Doppler shift, can be 
rewritten as an acceleration pointing towards the Earth (or Sun) of 
modulus 
\begin{equation}
  \label{eq:anomalous-acceleration}
  a_P = (8.74 \pm 1.33)\,10^{-10}\,{\rm m/s^2} \,.
\end{equation}
%


\newpage
\section*{\LARGE List of references}

\nociteEXPLAIN{Dvali.etal:2003}
\nociteEXPLAIN{Nieto.etal:2005}
\nociteEXPLAIN{Nottale:2003}
\nociteEXPLAIN{Palle:2005}
\nociteEXPLAIN{Ranada:2005}
\nociteEXPLAIN{Rosales.Sanchez-Gomez:1998}
\nociteEXPLAIN{Rosales:2002}
\nociteEXPLAIN{deDiego.etal:2005}
\nociteINFLUENCE{Anderson:1995}
\nociteINFLUENCE{Baker:2000}
\nociteINFLUENCE{Baker:2002}
\nociteINFLUENCE{Bel:2003}
\nociteINFLUENCE{Bolen.etal:2001}
\nociteINFLUENCE{Bonnor:1996}
\nociteINFLUENCE{Bonnor:1999}
\nociteINFLUENCE{Bonnor:2000a}
\nociteINFLUENCE{Celerier:2000}
\nociteINFLUENCE{Cooperstock.etal:1998}
\nociteINFLUENCE{Dicke.Peebles:1964}
\nociteINFLUENCE{Dirac:1979}
\nociteINFLUENCE{Dominguez.Gaite:2001}
\nociteINFLUENCE{Dumin:2002}
\nociteINFLUENCE{Dumin:2003}
\nociteINFLUENCE{Jaernefelt:1933}
\nociteINFLUENCE{Jaernefelt:1940}
\nociteINFLUENCE{Jaernefelt:1942}
\nociteINFLUENCE{Krasinsky.Brumberg:2004}
\nociteINFLUENCE{Mansouri:2005}
\nociteINFLUENCE{McVittie:1933}
\nociteINFLUENCE{Miller.Rouet:2002}
\nociteINFLUENCE{Mizony.Lachieze-Rey:2005}
\nociteINFLUENCE{Noerdlinger.Petrosian:1971}
\nociteINFLUENCE{Pachner:1963}
\nociteINFLUENCE{Pachner:1964}
\nociteINFLUENCE{Price:2005}
\nociteMATCHING{Balbinot.etal:1988}
\nociteMATCHING{Bonnor:1987}
\nociteMATCHING{Bonnor:2000b}
\nociteMATCHING{Burnett:1991}
\nociteMATCHING{Cahill.McVittie:1970a}
\nociteMATCHING{Cahill.McVittie:1970b}
\nociteMATCHING{Einstein.Straus:1945}
\nociteMATCHING{Einstein.Straus:1946}
\nociteMATCHING{Eisenstaedt:1975a}
\nociteMATCHING{Eisenstaedt:1975b}
\nociteMATCHING{Eisenstaedt:1977}
\nociteMATCHING{Ferraris.etal:1996}
\nociteMATCHING{Gautreau:1984a}
\nociteMATCHING{Gautreau:1984b}
\nociteMATCHING{Gautreau:2000}
\nociteMATCHING{Hawking:1968}
\nociteMATCHING{Hayward:1996}
\nociteMATCHING{Hayward:1998}
\nociteMATCHING{Hogan:1990}
\nociteMATCHING{Kantowski:1969}
\nociteMATCHING{Kantowski:1998}
\nociteMATCHING{Kramer.etal:2003}
\nociteMATCHING{Krasinski:1998}
\nociteMATCHING{Mars:1998}
\nociteMATCHING{Mars:2001}
\nociteMATCHING{Matravers.Humphreys:2001}
\nociteMATCHING{Mena.etal:2002}
\nociteMATCHING{Mena.etal:2003}
\nociteMATCHING{Mena.etal:2004}
\nociteMATCHING{Misner.Sharp:1964}
\nociteMATCHING{Nolan:1998}
\nociteMATCHING{Nolan:1999a}
\nociteMATCHING{Nolan:1999b}
\nociteMATCHING{Patel.etal:1999}
\nociteMATCHING{Schuecking:1954}
\nociteMATCHING{Senovilla.Vera:1997}
\nociteMATCHING{Sussman:1988}
\nociteMATCHING{vdBergh.Wils:1984}
\nociteMEASURE{Anderson.etal:1998}
\nociteMEASURE{Anderson.etal:2002a}
\nociteMEASURE{Anderson.etal:2002b}
\nociteMEASURE{Dittus.Laemmerzahl:2006}
\nociteMEASURE{Markwardt:2002}
\nociteMEASURE{Nieto.Turyshev:2004}
\nociteMEASURE{Turyshev.etal:2005a}
\nociteMEASURE{Turyshev.etal:2005b}
\nociteOTHERS{Bohm:2004}
\nociteOTHERS{Kinman:1992}
\nociteOTHERS{Lindegren.Dravins:2003}
\nociteOTHERS{Thornton.Border:2000}


\bibliographystyleMATCHING{siam}
\bibliographyMATCHING{references}

\bigskip

\bibliographystyleINFLUENCE{siam}
\bibliographyINFLUENCE{references}

\bigskip

\bibliographystyleEXPLAIN{siam}
\bibliographyEXPLAIN{references}

\bigskip

\bibliographystyleMEASURE{siam}
\bibliographyMEASURE{references}

\bigskip

\bibliographystyleOTHERS{siam}
\bibliographyOTHERS{references}


\end{document}